\documentclass[conferene,fleqn]{IEEEtran}
\IEEEoverridecommandlockouts
\IEEEpubid{\makebox[\columnwidth]{978-1-5386-5541-2/18/\$31.00~\copyright2018 IEEE \hfill} \hspace{\columnsep}\makebox[\columnwidth]{ }}
\usepackage{cite}
\usepackage{amsmath,amssymb,amsfonts}
\usepackage{algorithmic}
\usepackage{graphicx}
\usepackage{textcomp}
\usepackage{xcolor}
\usepackage[normalem]{ulem}
\usepackage{soul}
\usepackage{booktabs}
\usepackage{setspace}
\def\BibTeX{{\rm B\kern-.05em{\sc i\kern-.025em b}\kern-.08em
    T\kern-.1667em\lower.7ex\hbox{E}\kern-.125emX}}

\begin{document}


\title{SDN enabled Mobility Management in Multi Radio Access Technology 5G networks: A Survey}

\author{ \parbox{6 in}{\centering Pavan K Mangipudi \hspace{0.25in}
Janise McNair\\
\vspace{1mm}
Department of Electrical and Computer Engineering \\
University of Florida, Gainesville  USA\\
   \tt
   \small pavan.mangipudi@ufl.edu, mcnair@ece.ufl.edu
     }

}

\maketitle

\IEEEpubidadjcol

\begin{abstract}

This paper presents a survey of the state of the art in research related to handovers employing software defined networking (SDN) enabled architectures, serving multiple co-existing radio access technologies. As the industrial roll-out of cellular services continues to evolve, it brings with it the co-existence of various IP based networks such as 5G NR, LTE, WiFi, Satellite, and IoT networks. This coexistence of different radio access technologies presents researchers with the opportunity to use these technologies interchangeably to address the long-standing challenges associated with network and traffic management. Advances in software defined technologies including SDN enables handover and interoperability schemes that utilize the principles of SDN to improve the handover performance and achieve interconnection between these heterogeneous networks. This paper explores such advanced SDN enabled architectures for radio access networks, offloading techniques, and implementation approaches. Finally, the challenges and shortcomings of the SDN based handover optimization approaches are discussed, and a few future research paths are laid out.

\end{abstract}

\begin{IEEEkeywords}
SDN, 5G, WiFi, Multi RAT, Handovers, Mobility Management.
\end{IEEEkeywords}

\section{Introduction}
\label{intro}

5G technology has revolutionized the cellular and wireless industry. Its ultra-low latency, fast speeds, and bandwidths have made time-critical, real-time computing tasks possible. The cellular industry has expanded from just the sub 6 GHz frequency to millimeter wave (mm-Wave) frequencies, allowing for a substantial increase in subscribers and connected devices. While it took some time for the industry to start the roll-out of commercial 5G networks, it is now being done at a rapid rate across the world. Because of the expansion in cellular frequencies beyond the sub 6GHz range, there is now a co-existence of various IP based wireless networks like satellite, Internet of Things (IoT), Device to Device (D2D) networks, vehicular networks, etc. within or alongside the 5G networks. In conjunction with advancements in the cellular industry, another field that has gained interest from academia and industry alike is the domain of Software Defined Networking (SDN). SDN is a networking concept that segregates the control decisions of the network from forwarding hardware, by separating the network into control and data planes. Initially popular within wired networks, SDN has made its way to the wireless space. Organizations join the founding organization of SDN, Open Network Foundation (ONF) \cite{onf} to define controllers and protocols that make SDN compatible with wireless and cellular networks.

Advances in SDN-enabled handover schemes that utilize the principles of SDN to improve  handover performance and achieve interconnection between heterogeneous networks. In SDN, control functions are moved away from traditional switches to an external software entity, the SDN controller. The SDN controller monitors and defines the flow rules based on the traffic that is being sent and received in the switches. SDN networks are made up of OpenFlow enabled switches \cite{openflow1.2} which connect nodes with the SDN controller. These switches have multiple flow tables, which can be programmed through the SDN controller. Each incoming packet in the data plane is matched against all the flow tables in the switch, and when there is a match with a flow, the corresponding actions are executed by the switch. While the 5G architecture \cite{etsi5Gwithsdn} has a similar separation of data and control planes, the incorporation of SDN into cellular networks to the fullest of its capacities is yet to be seen. The applications of SDN in the standard are coupled with Network Function Virtualization to provide Network Slicing in 5G networks. So, the independent use of SDN in 5G networks to improve the handover is an emerging research area.

\begin{figure}[htbp!]
\centerline{\includegraphics[width=\columnwidth]{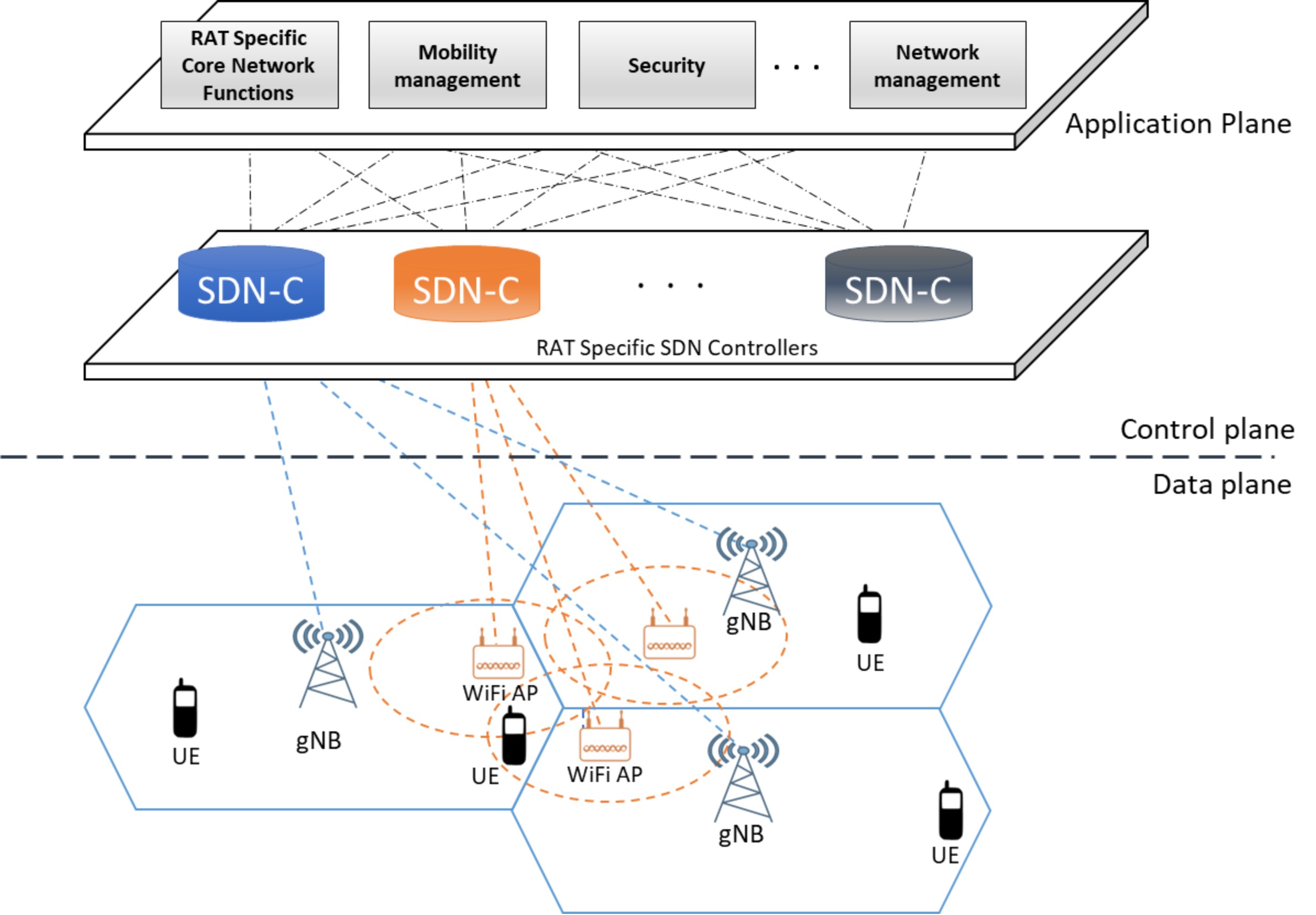}}
\caption{SDN based Multi RAT network topology.}
\label{MultiRAT_topo}
\end{figure}

\begin{figure}[htbp!]
\centerline{\includegraphics[width=\columnwidth]{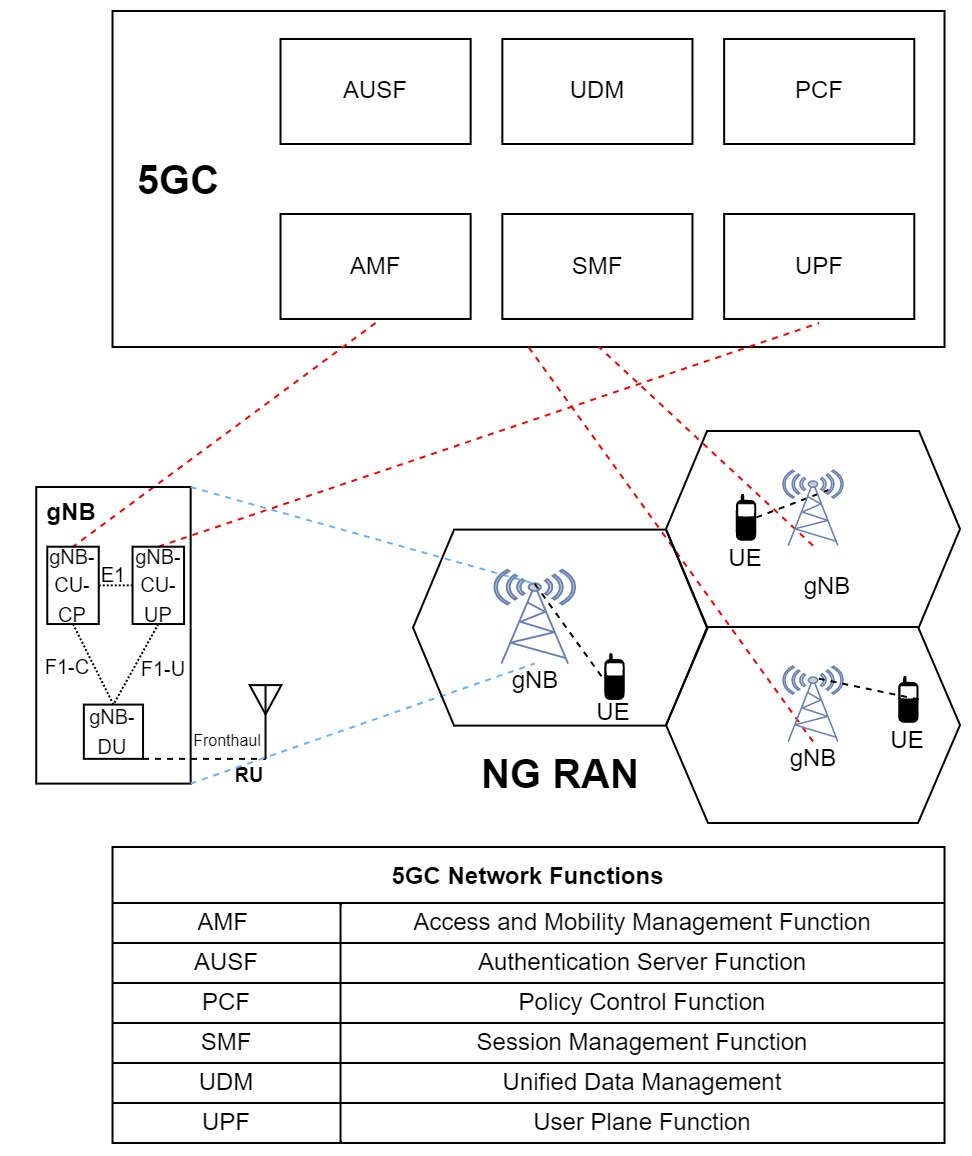}}
\caption{5G topology.}
\label{5G_topo}
\end{figure}

Another major challenge for mobility management in 5G and beyond networks is Multi Radio Access Network (RAT) mobility. Multi RAT networks are commonly found in the 5G networks, due to the coexistence among other networks for IoT, D2D, and M2M. Emerging research suggests that SDN can be utilized to leverage these Multi RAT networks to achieve interoperability between these co-existing networks. Fig.\ref{MultiRAT_topo} and Fig. \ref{5G_topo} show the network architectures of the current 5G network and the SDN based Multi RAT networks respectively.

There are several surveys that provide a comprehensive overview of the research on introducing SDN to legacy cellular networks. \cite{nguyen2017sdnnfvltesurvey} covers proposals to modify the LTE core network, the Evolved Packet Core(EPC) through SDN/NFV. It summarizes different ways in which SDN and NFV can be introduced into the EPC, covering the advantages and disadvantages of each approach. Similarly, \cite{nguyen2016sdnvirtualization} provides a comprehensive study of SDN and virtualization, and proposes a generalized SDN and virtualization based LTE network architecture for future mobile networks. It covers SDN and virtualization research in the RAN, backhaul, and core networks, using the LTE standards as the basis. \cite{haque2016wirelesssdnsurvey} is another survey that covers research on integrating SDN into cellular, mesh, sensor, and home networks. It explains how virtualization and a centralized approach using SDN can be integrated into the RAN and core architectures. These surveys, which were published before 5G was standardized, present a consolidated analysis of integrating SDN with the LTE architecture to make way for the then-upcoming 5G standards. They do not cover the research resulting from this integration, such as SDN based mobility and resource management, and handover procedures. Since 5G has been standardized, there is no survey that covers the emerging research on SDN based multi RAT 5G networks, which seeks to use SDN to address various problems associated with multi RAT networks. In this survey, we explore the architectural changes made to 5G networks to incorporate SDN and the resulting SDN based mobility management and handover techniques, which are not seen in other surveys. We investigate approaches to mobility management that are supported by or influenced by the new SDN-like architectures. In addition, we examine how SDN can provide a heterogeneous handover between co-existing networks with different RATs, which is not seen in other surveys. 

This paper is organized as follows: Section~\ref{background} provides definitions for the fundamental concepts, including SDN and handover in various wireless standards. Section \ref{ltesection} covers early SDN implementations using the 4G LTE architecture. Section \ref{5gsection} discusses approaches that modify the 3GPP's 5G NR architecture to deploy handover algorithms. In section \ref{multiratsection}, we explore the ability of SDN based handover techniques to manage multi RAT mobility. Section \ref{futurework} summarizes the limitations of the approaches seen in the survey and provides some directions which can be pursued in the future. Section \ref{conclusion} concludes this survey.

\begin{table*}[]
\centering
\caption{Categorization of papers used in this survey}
\label{organization_table}
\resizebox{\linewidth}{!}{%
\begin{tabular}{|l|lll|lll|lllll|}
\hline
\multicolumn{1}{|c|}{} &
  \multicolumn{3}{c|}{RAT type} &
  \multicolumn{3}{c|}{SDN Architecture type} &
  \multicolumn{5}{c|}{Handover optimization approach} \\ \hline
\multicolumn{1}{|c|}{} &
  \multicolumn{1}{c|}{LTE} &
  \multicolumn{1}{c|}{5G} &
  \multicolumn{1}{c|}{\begin{tabular}[c]{@{}c@{}}Multi \\ RAT\end{tabular}} &
  \multicolumn{1}{c|}{Centralized} &
  \multicolumn{1}{c|}{\begin{tabular}[c]{@{}c@{}}Semi \\ Centralized\end{tabular}} &
  \multicolumn{1}{c|}{Hierarchical} &
  \multicolumn{1}{c|}{\begin{tabular}[c]{@{}c@{}}Path \\ Prediction\end{tabular}} &
  \multicolumn{1}{c|}{\begin{tabular}[c]{@{}c@{}}SDN based\\ Handover\end{tabular}} &
  \multicolumn{1}{c|}{\begin{tabular}[c]{@{}c@{}}SDN based \\ Custom Interfaces\end{tabular}} &
  \multicolumn{1}{c|}{\begin{tabular}[c]{@{}c@{}}Multi Criteria \\ Decision Making\end{tabular}} &
  \multicolumn{1}{c|}{\begin{tabular}[c]{@{}c@{}}SDN based IP \\ Virtualization\end{tabular}} \\ \hline
Prados Et Al.\cite{prados2016handover} &
  \multicolumn{1}{l|}{\checkmark} &
  \multicolumn{1}{l|}{} &
   &
  \multicolumn{1}{l|}{\checkmark} &
  \multicolumn{1}{l|}{} &
   &
  \multicolumn{1}{l|}{} &
  \multicolumn{1}{l|}{\checkmark} &
  \multicolumn{1}{l|}{} &
  \multicolumn{1}{l|}{} &
   \\ \hline
Rizkallah Et.AL.\cite{rizkallah2018sdnverticalho} &
  \multicolumn{1}{l|}{\checkmark} &
  \multicolumn{1}{l|}{} &
   &
  \multicolumn{1}{l|}{} &
  \multicolumn{1}{l|}{\checkmark} &
   &
  \multicolumn{1}{l|}{} &
  \multicolumn{1}{l|}{} &
  \multicolumn{1}{l|}{} &
  \multicolumn{1}{l|}{\checkmark} &
   \\ \hline
Gharsallah Et.Al.\cite{gharsallah2019sdn} &
  \multicolumn{1}{l|}{\checkmark} &
  \multicolumn{1}{l|}{} &
   &
  \multicolumn{1}{l|}{\checkmark} &
  \multicolumn{1}{l|}{} &
   &
  \multicolumn{1}{l|}{\checkmark} &
  \multicolumn{1}{l|}{} &
  \multicolumn{1}{l|}{} &
  \multicolumn{1}{l|}{} &
   \\ \hline
Lee Et. Al.\cite{lee2017handover4g} &
  \multicolumn{1}{l|}{\checkmark} &
  \multicolumn{1}{l|}{} &
   &
  \multicolumn{1}{l|}{\checkmark} &
  \multicolumn{1}{l|}{} &
   &
  \multicolumn{1}{l|}{\checkmark} &
  \multicolumn{1}{l|}{} &
  \multicolumn{1}{l|}{} &
  \multicolumn{1}{l|}{} &
   \\ \hline
Abdulghaffar Et Al.\cite{abdulghaffar2021modeling} &
  \multicolumn{1}{l|}{} &
  \multicolumn{1}{l|}{\checkmark} &
   &
  \multicolumn{1}{l|}{\checkmark} &
  \multicolumn{1}{l|}{} &
   &
  \multicolumn{1}{l|}{} &
  \multicolumn{1}{l|}{\checkmark} &
  \multicolumn{1}{l|}{} &
  \multicolumn{1}{l|}{} &
   \\ \hline
Dhruvik Et Al.\cite{dhruvik2021design} &
  \multicolumn{1}{l|}{} &
  \multicolumn{1}{l|}{\checkmark} &
   &
  \multicolumn{1}{l|}{\checkmark} &
  \multicolumn{1}{l|}{} &
   &
  \multicolumn{1}{l|}{} &
  \multicolumn{1}{l|}{\checkmark} &
  \multicolumn{1}{l|}{} &
  \multicolumn{1}{l|}{} &
   \\ \hline
Ciciouglu Et Al.\cite{ciciouglu2021multi} &
  \multicolumn{1}{l|}{} &
  \multicolumn{1}{l|}{\checkmark} &
   &
  \multicolumn{1}{l|}{\checkmark} &
  \multicolumn{1}{l|}{} &
   &
  \multicolumn{1}{l|}{} &
  \multicolumn{1}{l|}{} &
  \multicolumn{1}{l|}{} &
  \multicolumn{1}{l|}{\checkmark} &
   \\ \hline
Taksande Et Al.\cite{taksande2020open5g} &
  \multicolumn{1}{l|}{} &
  \multicolumn{1}{l|}{} &
  \checkmark &
  \multicolumn{1}{l|}{\checkmark} &
  \multicolumn{1}{l|}{} &
   &
  \multicolumn{1}{l|}{} &
  \multicolumn{1}{l|}{} &
  \multicolumn{1}{l|}{\checkmark} &
  \multicolumn{1}{l|}{} &
   \\ \hline
Wang Et Al.(2016)\cite{wang2016convergence} &
  \multicolumn{1}{l|}{} &
  \multicolumn{1}{l|}{} &
  \checkmark &
  \multicolumn{1}{l|}{\checkmark} &
  \multicolumn{1}{l|}{} &
   &
  \multicolumn{1}{l|}{} &
  \multicolumn{1}{l|}{} &
  \multicolumn{1}{l|}{} &
  \multicolumn{1}{l|}{} &
   \\ \hline
Alfoudi Et Al.\cite{Alfoudi2019seamless} &
  \multicolumn{1}{l|}{} &
  \multicolumn{1}{l|}{} &
  \checkmark &
  \multicolumn{1}{l|}{} &
  \multicolumn{1}{l|}{} &
  \checkmark &
  \multicolumn{1}{l|}{} &
  \multicolumn{1}{l|}{\checkmark} &
  \multicolumn{1}{l|}{} &
  \multicolumn{1}{l|}{} &
   \\ \hline
Alotaibi Et Al. (2018)\cite{alotaibi2018hierarchical} &
  \multicolumn{1}{l|}{} &
  \multicolumn{1}{l|}{} &
  \checkmark &
  \multicolumn{1}{l|}{} &
  \multicolumn{1}{l|}{} &
  \checkmark &
  \multicolumn{1}{l|}{\checkmark} &
  \multicolumn{1}{l|}{} &
  \multicolumn{1}{l|}{} &
  \multicolumn{1}{l|}{} &
   \\ \hline
Alotaibi Et Al. (2021)\cite{alotaibi2021linking} &
  \multicolumn{1}{l|}{} &
  \multicolumn{1}{l|}{} &
  \checkmark &
  \multicolumn{1}{l|}{} &
  \multicolumn{1}{l|}{} &
  \checkmark &
  \multicolumn{1}{l|}{} &
  \multicolumn{1}{l|}{} &
  \multicolumn{1}{l|}{} &
  \multicolumn{1}{l|}{\checkmark} &
   \\ \hline
Liyanage Et Al.\cite{liyanage2017sdnoffloading} &
  \multicolumn{1}{l|}{} &
  \multicolumn{1}{l|}{} &
  \checkmark &
  \multicolumn{1}{l|}{} &
  \multicolumn{1}{l|}{\checkmark} &
   &
  \multicolumn{1}{l|}{} &
  \multicolumn{1}{l|}{\checkmark} &
  \multicolumn{1}{l|}{} &
  \multicolumn{1}{l|}{} &
   \\ \hline
Wang Et Al.(2017)\cite{wang2017sdn} &
  \multicolumn{1}{l|}{} &
  \multicolumn{1}{l|}{} &
  \checkmark &
  \multicolumn{1}{l|}{\checkmark} &
  \multicolumn{1}{l|}{} &
   &
  \multicolumn{1}{l|}{} &
  \multicolumn{1}{l|}{} &
  \multicolumn{1}{l|}{} &
  \multicolumn{1}{l|}{} &
  \checkmark \\ \hline
\end{tabular}%
}
\end{table*}


\begin{table}[]
\caption{\label{acronyms}List of Abbreviations used in this paper}

\resizebox{\columnwidth}{!}{%
\begin{tabular}{c|c}
\hline
Abbreviation & Meaning                                                               \\ \hline\hline
3GPP         & 3\textsuperscript{rd} Generation Partnership Project \\ \hline
4G LTE       & 4\textsuperscript{th} Generation Long Term Evolution \\ \hline
5G NR        & 5\textsuperscript{th} Generation New Radio           \\ \hline
5GC          & 5G Core Network                                                       \\ \hline
AMF          & Access and Mobility Management Function                               \\ \hline
AP           & Access Point                                                          \\ \hline
BLER         & Block Error Rate                                                      \\ \hline
BS           & Base Station                                                          \\ \hline
CN           & Core Network                                                          \\ \hline
D2D          & Device to Device Communication                                        \\ \hline
eNB          & Evolved Node B                                                        \\ \hline
EPC          & Evolved Packet Core (4G)                                              \\ \hline
gNB          & Next generation Node B                                                \\ \hline
GPRS         & General Packet Radio Service                                          \\ \hline
GPS          & Global Positioning System                                             \\ \hline
GTP-U        & GPRS Tunnelling Protocol – User                                       \\ \hline
HetNet       & Heterogenous Network                                                  \\ \hline
IoT          & Internet of Things                                                    \\ \hline
IP           & Internet Protocol                                                     \\ \hline
LiFi         & Light Fidelity technology                                             \\ \hline
LWA          & LTE WiFi Aggregation                                                  \\ \hline
LoRaWAN      & Long Range Wide Area Network                                          \\ \hline
MAC          & Medium Access Control                                                 \\ \hline
MPLS         & Multi Protocol Label Switching                                        \\ \hline
M2M          & Machine to Machine Communication                                      \\ \hline
mmWave       & Milli-meter Wave                                                       \\ \hline
MME          & Mobility Management Entity                                            \\ \hline
NFV          & Network Function Virtualization                                       \\ \hline
NGRAN        & Next Generation Radio Access Network of 5G NR                         \\ \hline
OF           & OpenFlow                                                              \\ \hline
ONF          & Open Network Foundation                                               \\ \hline
PDN-GW/PGW   & Packet Data Network Gateway                                           \\ \hline
PDU          & Protocol Data Unit Session                                            \\ \hline
PHY          & Physical Layer                                                        \\ \hline
RAT          & Radio Access Technology                                               \\ \hline
RAN          & Radio Access Network                                                  \\ \hline
RLC          & Radio link control                                                    \\ \hline
RSS          & Received Signal Strength                                              \\ \hline
RRC          & Radio Resource Control                                                \\ \hline
RSSI         & Receiver Signal Strength Index                                        \\ \hline
SINR         & Signal to Interference and Noise Ratio                                \\ \hline
SMF          & Session Management Function                                           \\ \hline
SDN          & Software Defined Networking                                           \\ \hline
S-GW/SGW     & Service Gateway                                                       \\ \hline
TCP          & Transmission Control Protocol                                         \\ \hline
UPF          & User Plane Function                                                   \\ \hline
UE           & User Equipment                                                        \\ \hline
WiFi         & Wireless Fidelity                                                     \\ \hline
WLAN         & Wireless Local Area Network                                           \\ \hline
\end{tabular}%
}
\end{table}

\section{Background}
\label{background}

Cellular networks constitute the multiple generations of mobile phone communication ranging from the 1\textsuperscript{st} generation's analog communications to the current 5\textsuperscript{th} generation's 5G new radio. As the name suggests, the communication is spread out over service areas defined as cells, with each cell assumed to take a hexagonal shape. Cellular networks can be widely divided into two parts, the core network (CN) and the radio access network (RAN). The RAN is part of a mobile network that connects user equipment to the core network. In each cell is a tower or Base Station (BS) which connects with the mobile user equipment (UE) through the air interface, and provides cellular coverage to the UEs. Cellular network sizes range from a few kilometers (macro cells) to a few meters (femtocells), depending on the BS's transmission power and frequency of transmission. The movement of users across adjacent cells necessitates the UEs to reconnect to the other cell's base station so that the connection is not lost. This involves the transfer of the data coming from the core network between the two base stations, and the re-establishment of the connection between UE and BS. This phenomenon of transfer of a UE from one cell to another is called handover.

\subsection{Handover in cellular networks}

\begin{figure}[t]
\centerline{\includegraphics[width=\columnwidth]{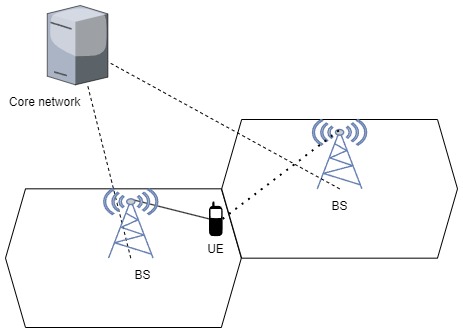}}
\caption{Simple cellular network topology during a handover.}
\label{handover}
\end{figure}

 Fig.\ref{handover} shows the topology of a simple cellular network with the UE moving from one cell to another. Handover can occur in cellular networks due to different reasons. These include, but are not restricted to movement of users across different cells, the transfer of a user from one channel to another to avoid interference, or the transfer of users between smaller and larger overlapping cells when capacity in one of the cells is approaching its limit. Any handover operation is a three-stage process that includes handover decision, which includes data collection and analysis for handover detection and the handover trigger; radio link transfer, which includes flow and path determination and routing,
and channel assignment, which includes the radio frequency channel selection~\cite{mcnairvertho2004}. This paper discusses the handover decision and radio link transfer components.  

\begin{figure}[t]
\centerline{\includegraphics[width=\columnwidth]{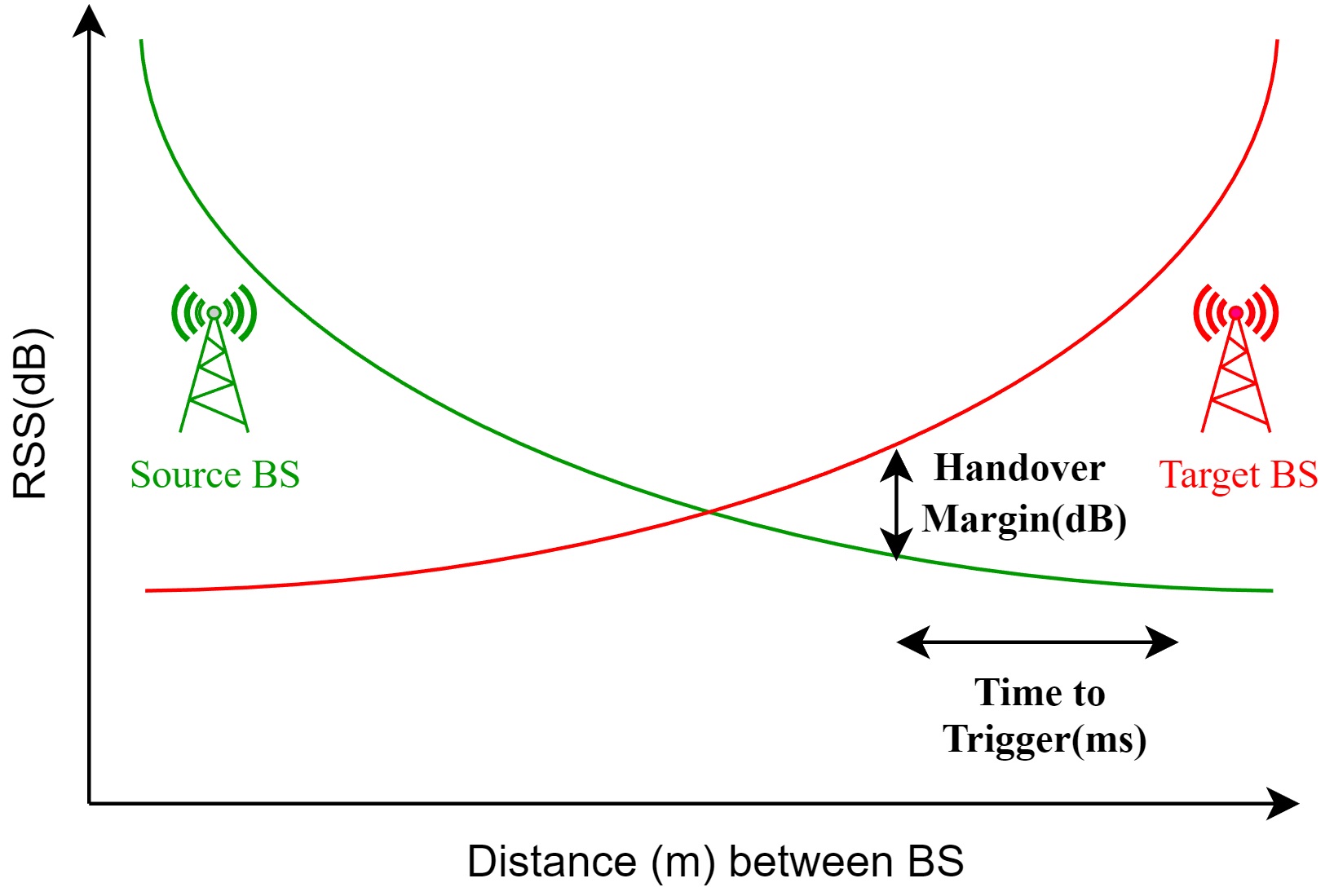}}
\caption{Hysteresis based general handover.}
\label{hysteresis}
\end{figure}

The cell the UE is currently in is referred to as the source and the cell the UE intends to move to is called the target cell. Handover is triggered by the source BS when the received signal strength (RSS) of the UE at the source BS falls below the RSS of the UE at the target base station. Different generations of cellular communication define the triggering of handover in different ways. A simple description of the handover triggering event is shown in Fig.\ref{hysteresis}. The handover margin is the range, in dB, beyond which the UE's connection with the source BS is lost. So, Handover is triggered once the RSS at the source reaches the Handover Margin level. This is a general understanding of the HO trigger. There exist other parameters such as time to trigger(TTT) which define how long the source BS waits after Handover Margin is reached to trigger the handover. This is used to make sure the UE's RSS does not go back and forth due to movement between the cells.

In cellular networks, handovers can also be classified into Horizontal and Vertical handovers, based on the RATs involved in the handover. The horizontal handover refers to the handover between cells of the same RAT. Vertical handover refers to the handover between cells belonging to different RATs. Vertical RATs are becoming increasingly common due to the existence of legacy RATs such as LTE and 3G within the newer 5G networks. Additionally, Vertical handover is also seen between cellular and WiFi RATs. LTE-WiFi aggregation (LWA) \cite{lwalwip3gpp} is an example of a vertical handover. Another classification of handovers is homogeneous and heterogeneous handover. Homogeneous handover, as the name suggests, is the case pertaining to the user moving between APs belonging to the same access network type, i.e. 5g to 5g or WiFi to WiFi. Heterogeneous handover refers to the case of the user moving between APs of different access network types, such as between LTE and WiFi. 

\subsection{Xn based Handover}
In 5G NR, the algorithm used for handover is the Xn based handover. Xn handover is based on an evolution of the LTE X2 handover. The differences between the X2 and Xn based handovers consist of the interfaces used and Xn based handover supports handover between different frequencies. For this reason, we first describe the X2 based handover, followed by the updates in 5G.

\subsection{X2 based handover in 4G LTE}
\label{x2hobg}
This section covers X2 based handover, defined in the 4G LTE standard \cite{etsiltex2ho}. X2 is the interface between the UE's and the BS, called the evolved Node-B (eNB). The handover is managed completely by the eNBs, without interruption from the core network. 

The main components of the LTE architecture that are involved in the X2 Handover are the eNB, Service Gateway (S-GW), Packet Data Network Gateway (PDN-GW), and the Mobility Management Entity (MME). The MME, S-GW, and PDN-GW are a part of the LTE core network, called the Evolved Packet Core (EPC). 

The functionalities of these components are as follows:

\begin{itemize}
    \item eNodeB(eNB) – or the Base Station, contains the radio channel and all IP-level connectivity to the MME/S-GW/PDN-GW to create and maintain connectivity between the radio channel and the core network.
    \item S-GW: Refers to a component that connects the eNB with the PDN-GW and is usually a router that sends the tunnel to the PDN-GW with data from the eNB or the UE.
    \item PDN-GW: Functions as the anchor or the home of the network, and is the main Network Address Translation (NAT)/Routing GW to the internet. The PDN-GW keeps track of all tunnels for each UE on the network and knows how to forward packets to the UEs. 
    \item MME: The Mobility Management Entity, handles mobility in general and handles many mobility functions coordinating with the PDN-GW/S-GWs.
\end{itemize}

\begin{figure}[htbp]
\centerline{\includegraphics[width=9cm]{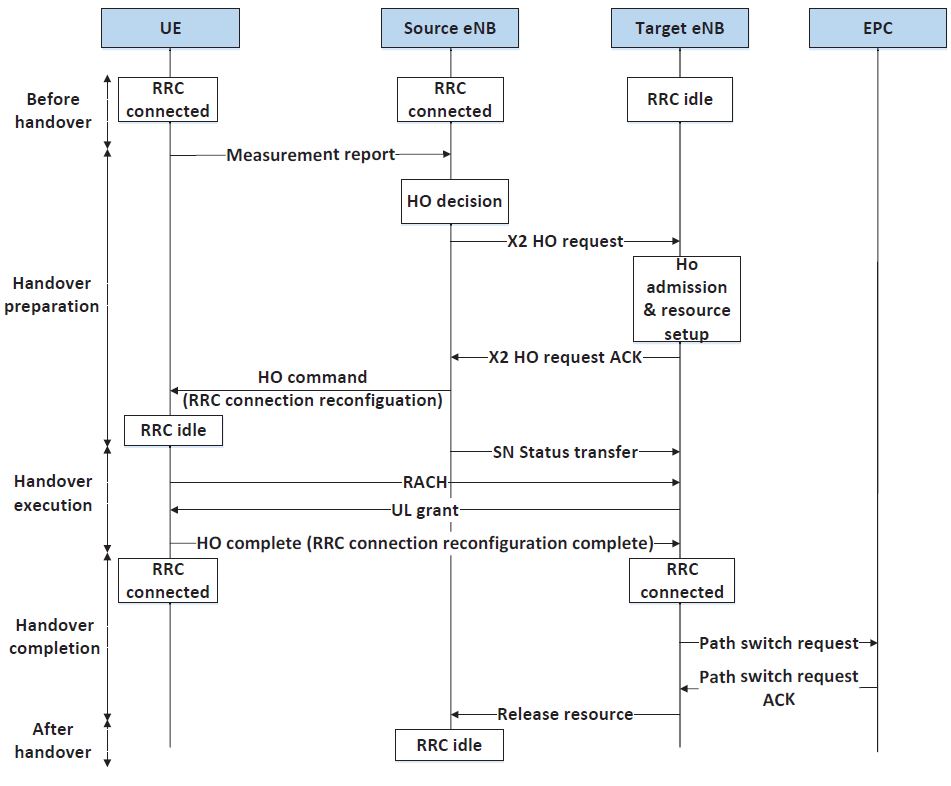}}
\caption{X2 based handover in LTE, as explained in \cite{x2hopaper}.}
\label{ltex2ho}
\end{figure}

X2 handover of LTE is completely performed between the two eNBs involved. The MME is involved after the completion of the handover procedure to initiate the path switch procedure. The detailed X2 handover call flow is described in Fig.\ref{ltex2ho}.

\subsubsection{Handover Decision}
The UE periodically sends Radio Resource Control (RRC) Measurement Reports (MRM) to the source eNB (base station) about neighboring cells. Handover is triggered depending on measurement results, and if one of the handover trigger events described in the standard is satisfied. There are a total of 8 measurement events defined, depending on the source eNB's signal levels and target eNB's signal levels. In such an event, the source eNB decides X2 handover and sends an X2 Handover Request to the target eNB. This message contains information needed to perform the handover such as UE context information, Target cell ID, and Radio access bearer (RAB) context. If the target eNB is able to provide the required quality of service for the new UE, it sends a handover acknowledgment (ACK) to the source through the X2 direct tunnel. Additionally, the target eNB also establishes an uplink S1 bearer with the same S-GW with which the source eNB has been connected. If the target eNB cannot accept the handover request, an X2 failure message is sent to the source eNB. The tunnels are created such that all internet traffic to the UE will start flowing to the target eNB.

The source eNB receives the ACK, which also contains the configuration of the General Packet Radio Service (GPRS) Tunnelling Protocol (GTP-U) tunnels as well as the RRC Connection Reconfiguration message that needs to be forwarded to the UE. The RRC message consists of Layer1/Layer2 (first two layers of the LTE protocol stack) parameters for the UE, which are needed for synchronization with the target eNB. Finally, the handover command is sent to the UE by the source eNB, with the RRC connection Reconfiguration message obtained in the previous step. 

\subsubsection{Handover Radio Link Transfer}

The UE receives the RRC reconfiguration message and shifts into the RRC idle state, where the connection with the source eNB is severed. Upon detachment, the source eNB sends a Sequence Number (SN) status transfer message containing the Packet Data Convergence Protocol (PDCP) sequence numbers to the target eNB. This is done so that the packets that are received from the internet to the UE can be synchronized using these PDCP sequence numbers. Next, the UE is synchronized with the target eNB using the RRC reconfiguration message it received, completing the successful handover. After the completion of the handover, the target eNB forwards the packets received from the source eNB during the handover process to the UE. So, the buffered data is transferred to the UE from the target eNB. A path switch procedure is initiated between the target eNB and the MME to update the position of the UE. The S-GW establishes a downlink S1 bearer with the target eNB and the S-GW switches the data path from the source eNB to the target eNB and releases the old S1 bearer. This concludes the handover procedure.

\subsection{Xn based Handover in 5G}

\begin{figure}[htbp]
\centerline{\includegraphics[width=\columnwidth]{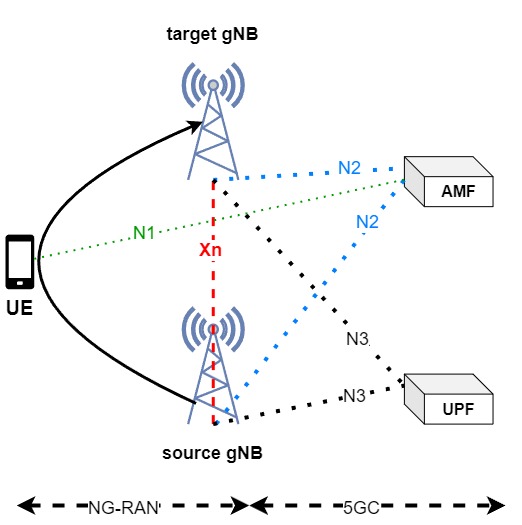}}
\caption{Block diagram of the 5G Xn handover with interfaces.}
\label{5GXnHOdiagram}
\end{figure}

In 5G NR, the algorithm used for handover is the Xn based handover. This is very similar to the X2 handover seen in LTE. Fig.\ref{5GXnHOdiagram} shows the interfaces involved between the UE and the 5G base station called as gNB. For this architecture, the next-generation RAN (NG-RAN) and 5G Core network (5GC) are defined. Fig.\ref{5GXnHOdiagram} also shows the core network components Access and Mobility Management Function (AMF) and User Plane Function (UPF), which are involved in the Xn handover, similar to the case of the X2 handover. 

In the transition from LTE to 5G, the MME's functionality is split between the AMF and Session Management Function (SMF). The AMF receives all the connection and session related information but is only responsible for handling the connection and mobility management functionalities. Session management messages it receives are forwarded to the SMF over the N11 interface.

\begin{figure*}[t!]
\centerline{\includegraphics[width=\textwidth]{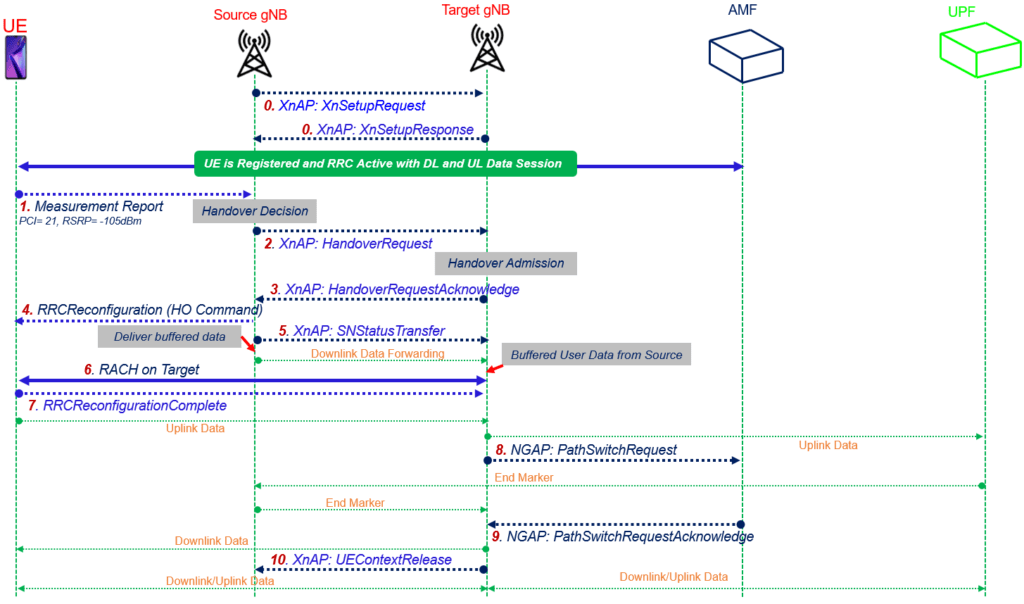}}
\caption{Call flow of the Xn based 5G Handover, without UPF reallocation.}
\label{5GXnCallFlow}
\end{figure*}

There are minimal differences between the X2 and Xn based handovers. The interface used is the Xn interface of 5G instead of the X2 in LTE. Xn based handover supports handover between different frequencies. These are Inter Frequency and Intra frequency handover. MME, S-GW, PDN-GW in case of LTE are replaced by the AMF and UPF in the 5G handover. Two types of handovers are seen in 5G using the Xn handover. Inter gNB and Intra gNB handover. In the inter-gNB handover, the UE moves between gNBs that are in the coverage area of the same UPF. So, UPF reallocation is not necessary. Intra gNB handover is the case where the UE moves between gNBs belonging to different UPF service areas This requires reallocation of UPF.  Fig.\ref{5GXnCallFlow} shows the call flow for an Inter gNB Xn based handover, as seen in \cite{xnhosite}. 

From Fig.\ref{5GXnCallFlow}, it can be seen that the handover steps are very similar to the ones described in Section \ref{x2hobg}. The change is in the interfaces and the core network elements. However, the signaling involved in the handover decision and radio link transfer is similar to the X2 handover. Another technology bridging point is that the 5G architecture \cite{etsi5Gwithsdn} has a similar separation of data and control planes as found in software defined networks (SDNs). In the next section, we provide background on SDN.



\subsection{Software Defined Networks (SDN) and Handover Management}

\begin{figure}[htbp!]
\centerline{\includegraphics[width=\columnwidth]{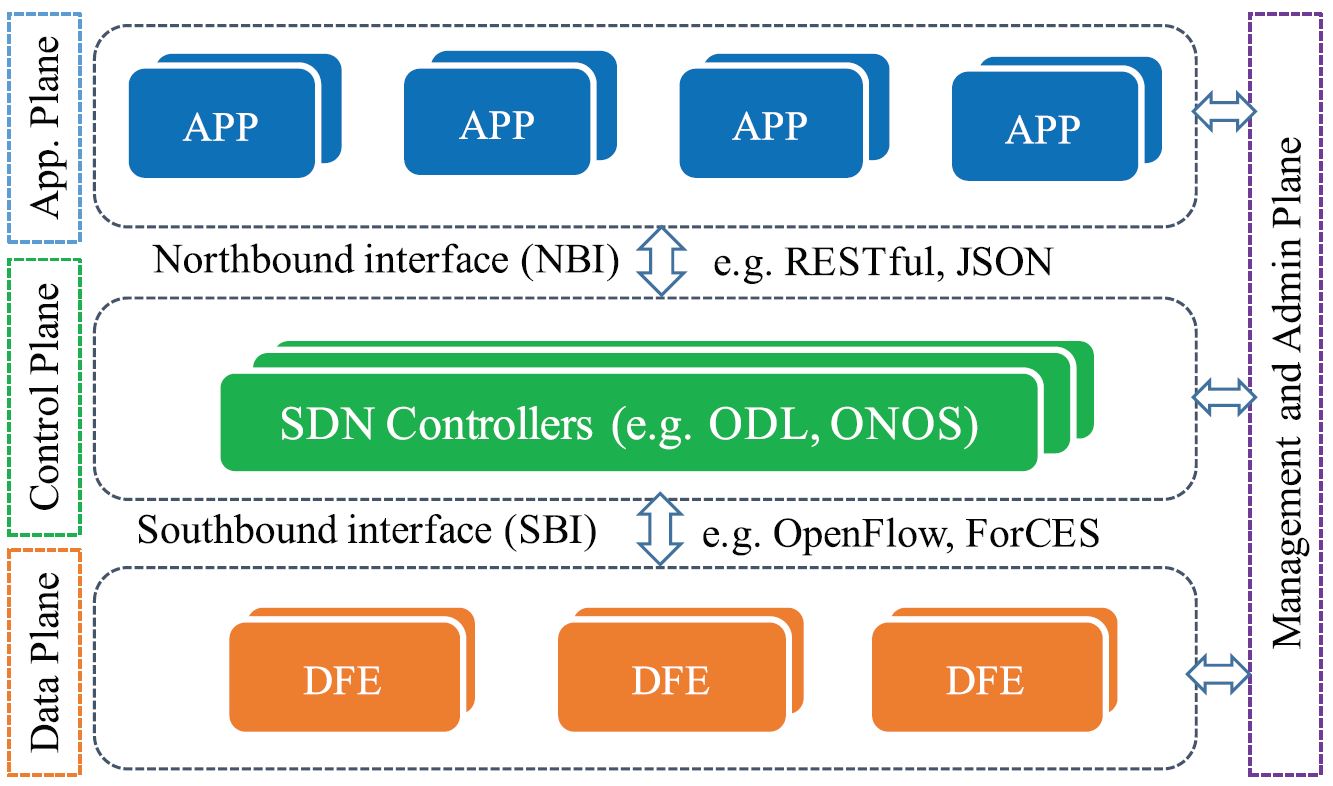}}
\caption{A high level SDN architecture \cite{sdnnfvsurvey}.}
\label{SDNarch}
\end{figure}

\begin{figure}[htbp]
\centerline{\includegraphics[width=\columnwidth]{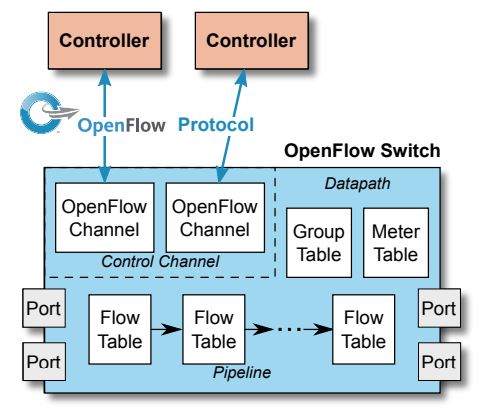}}
\caption{OpenFlow Logical Switch \cite{ovsv1.5.1}.}
\label{openflowSwitch}
\end{figure}

As mentioned in the Introduction, SDN is a  network management paradigm, which separates the decisions of the control plane (packet handling and routing) from the data plane (operations that forward the packets in a network )\cite{wirelesssdnsurvey}. Fig.\ref{SDNarch} gives a high-level overview of the architecture of an SDN. The control plane of the network includes one or more controllers, which have a global view of the network, to carry out the network management and control functions. The application layer is on top of the control layer and consists of the various applications which decide the policies administered through the controllers, such as data routing, load balancing, etc. These applications communicate their network requirements to the controllers through the northbound interfaces (NBI). Below the control plane, the data plane consists of the data forwarding entities such as physical/virtual switches and routers which are capable of communicating with the SDN controller through the southbound interface. The management and administration plane is responsible for functions such as provisioning and monitoring of the networks. The first protocol to be developed for SDN, the OpenFlow protocol\cite{openflow}, is maintained by the Open Networking Foundation (ONF)\cite{onf}. In OpenFlow enabled switches, shown in Fig. \ref{openflowSwitch}, the flow tables and a group table are responsible for the packet lookup and forwarding, and the OpenFlow channels connect to the SDN Controller. A flow in SDN is a set of instructions transmitted from the SDN Controller to the switch. Each individual flow contains packet match fields, flow priority, various counters, packet processing instructions, flow timeouts, and a cookie. The flows are organized in tables, called flow tables. Using the OpenFlow protocol, the SDN Controller can add, update, and delete flows or flow entries in the flow tables, in a reactive or proactive manner.  The controller uses the global view of the network to make efficient and intelligent routing decisions and sends the decisions to the user plane through the flow tables in the OpenFlow switches. 

The controller in SDN can be implemented in three ways. Centralized, Distributed, and Hierarchical are the three main implementations. A centralized approach has a single controller that controls the network. The limitation is the single controller may become a bottleneck or a single point of failure as the traffic across the network increases. To overcome this, multiple controllers are used, in either a distributed/flat or hierarchical manner. In a Distributed or Flat controller approach, multiple controllers operate on their local view. A benefit of this approach is that by dividing the role between the controllers of different layers heavy loads are distributed between the controllers, reducing the control signal overhead. Synchronization is needed to get a global view of the network. This approach provides increased resilience to failure but also increases overhead through an increase in the control messages exchanged between the controllers. A Hierarchical architecture, on the other hand, uses multiple levels of controllers, generally two or three. The lower layer of controllers is responsible for directly interacting with the data plane, while the higher layers manage the controllers in the layer below it. The advantage of using this approach is that the management of the controllers is simplified. A hierarchical controller setup covers the drawbacks of centralized and distributed approaches. It reduces control signaling overhead as it allows for better aggregation of control information. It also solves the scalability issue of the centralized approach as a central controller doesn't have to handle traffic growth over a large area. However, if a single controller is designated at the top level, then the single point of failure remains \cite{sdnarchcontrollersurvey}.



\section{Early SDN implementations based on LTE Architectures}
\label{ltesection}

\begin{figure}[htbp]
\centerline{\includegraphics[width=\columnwidth]{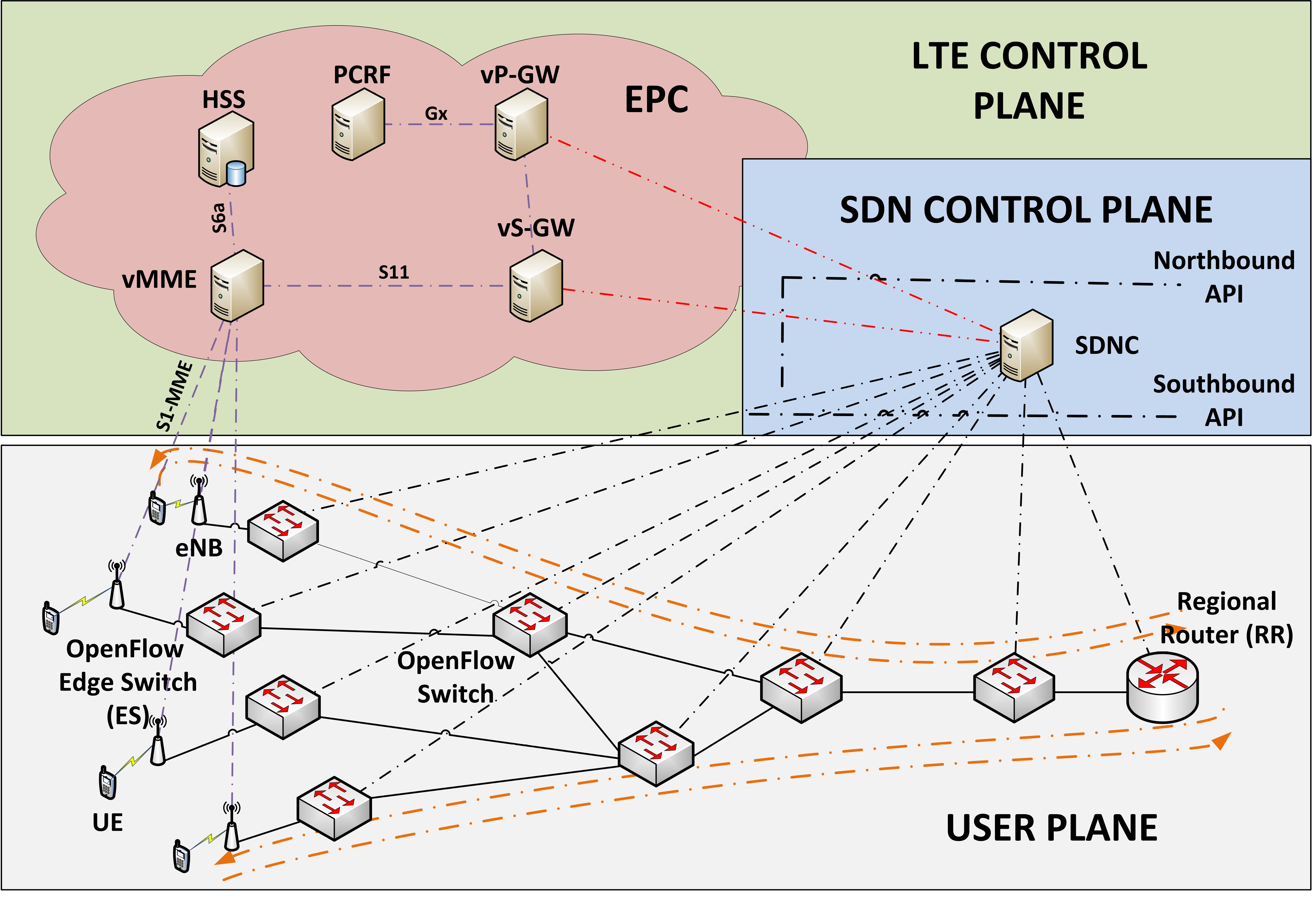}}
\caption{Proposed Architecture of \cite{prados2016handover}.}
\label{arch2}
\end{figure}

\begin{figure}[htbp]
\centerline{\includegraphics[width=\columnwidth]{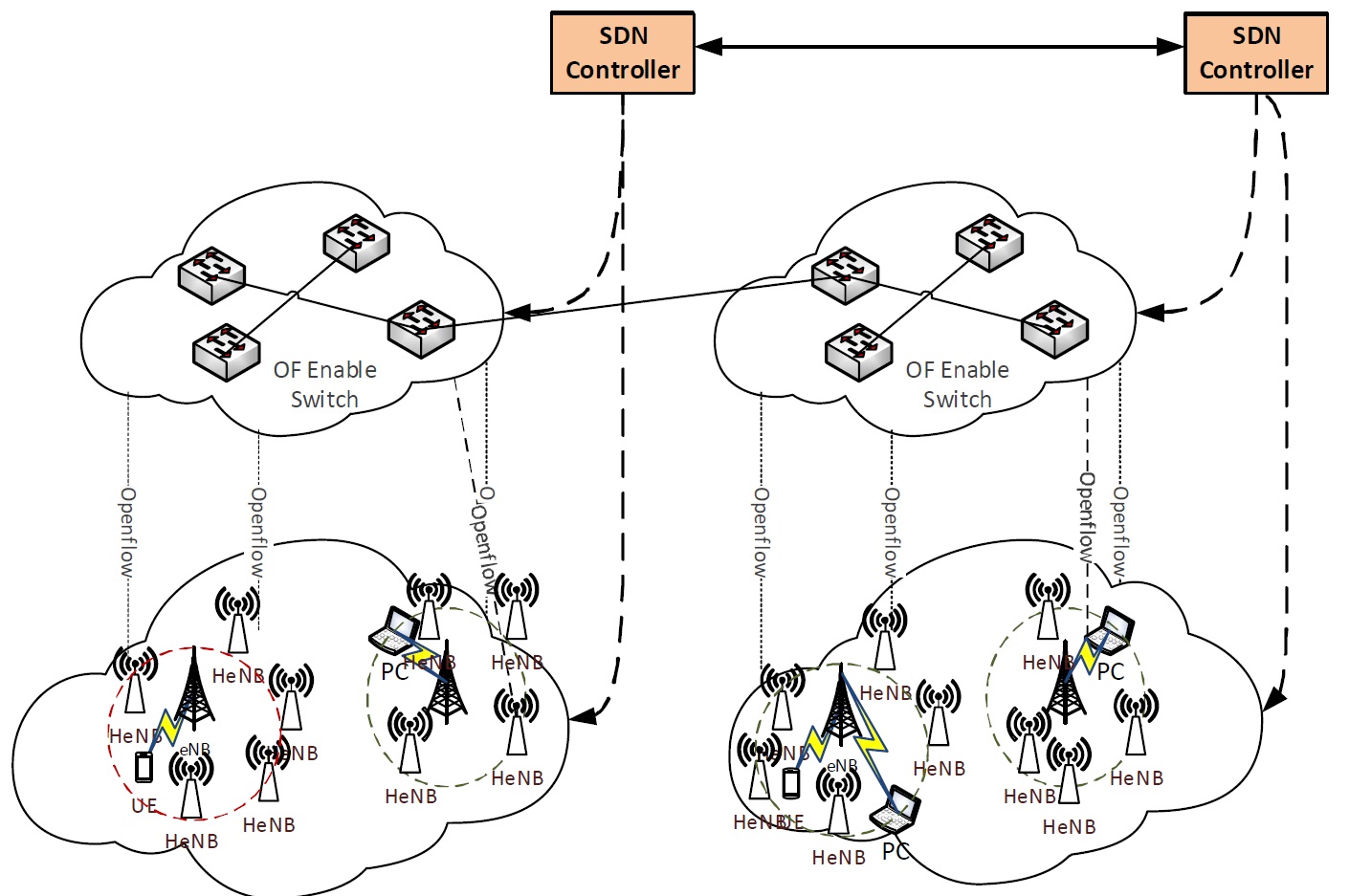}}
\caption{Network Topology of \cite{rizkallah2018sdnverticalho}.}
\label{arch1}
\end{figure}

While recent research moves towards technically sophisticated algorithms addressing very domain-specific problems, the early research in this domain focused on modifying the existing LTE architecture to incorporate SDN techniques \cite{prados2016handover}, \cite{rizkallah2018sdnverticalho}, \cite{gharsallah2019sdn}, \cite{lee2017handover4g}. Although this paper was published before 3GPP’s specifications for 5G were finalized, \cite{prados2016handover} serves as a good starting point in understanding how SDN can be integrated into cellular architectures. The central idea of these works is to leverage the global view of the network provided by the SDN controller and to develop efficient mobile routing, handovers, and cellular routing policies that reduce control message exchanges involved in these processes. Specifically, in \cite{prados2016handover} and \cite{gharsallah2019sdn}, the SDN is introduced, in conjunction with Network Function Virtualization (NFV), in a centralized manner where the SDN Controller has the view of the entire network (e.g. Fig.\ref{arch2}). \cite{prados2016handover} proposes SDN coupled with NFV to optimize the X2 based handover\cite{etsiltex2ho}, while \cite{gharsallah2019sdn} proposes an SDN and NFV based architecture with a software defined Handover Management Entity. Along with the virtualized network functions of the core network, the handover management entity prepares virtualized cells and anchor points to aid in the optimization of the handover in ultra dense networks. \cite{rizkallah2018sdnverticalho} uses an SDN based semi-centralized architecture to propose a novel vertical handover decision scheme for handover from a macro cell to femtocells in a 5G HetNet (Fig.\ref{arch1}). \cite{lee2017handover4g} does not specify its SDN architecture. However, SDN is proposed to be used to implement a linear programming algorithm for seamless handover.

\subsection{Centralized SDN with NFV to Optimize LTE X2 handover} 

\subsubsection{Architecture}
The centralized architecture proposed in \cite{prados2016handover} uses SDN and NFV where the LTE control plane entities (i.e., P-GW, S-GW) are implemented as virtual functions in a data center, and the User Plane is made up of OpenFlow enabled switches distributed across the network to  connect the User Plane entities such as the evolved Node-B (eNBs) and user equipment (UE) to the SDN Controller. This way, the SDN Controller acts as the intermediary between the control plane comprising of the virtualized Control Plane entities and the User Plane switches and eNBs, realizing the dis-aggregation of the control and data planes, and providing the aforementioned global view of the network. The X2 based handover of LTE is modified by replacing the GTP-U protocol with multi protocol label switching (MPLS) tunnels, which are handled by the SDN Controller, resulting in a significant reduction of overhead in each User Plane packet.

 Using this architecture, the authors of \cite{prados2016handover} propose a modified X2 handover that involves the following changes to the X2 interface:
\begin{itemize}
    \item Every eNB stores a Neighbour Information Table (NIT) which interrelates the Physical Cell Identifier (PCI), Evolved Cell Global Identifier, and the IP address of its neighboring eNBs. 
    \item eNBs discover their neighboring eNBs through the LTE Automatic Neighbour Relation Function or by directly requesting the SDN Controller.
    \item The SDN Controller stores a table called Network Information Base, which contains all the identifiers allocated for all of the network entities (e.g., Physical Cell ID, IP addresses, MPLS labels, etc.).
\end{itemize}

\subsubsection{Handover Decision Process}

The information used in the handover decision of \cite{prados2016handover} is exactly the same as that used in an X2 based handover. The situation considered is that of the UE being connected to the network, and the network knowing the UE's location precisely through the RSSI, PCI, and IP addresses reported to it by the UE. RSSI is the only factor considered for the handover, and the decision to do a handover is triggered when the RSSI of the target eNB becomes greater than the RSSI at the source eNB. The proposed OpenFlow based X2 handover is as shown in Fig. \ref{ho2procedure}.

\subsubsection{Handover Radio Link Transfer Process}

\begin{figure}[htbp]
\centerline{\includegraphics[width=\columnwidth]{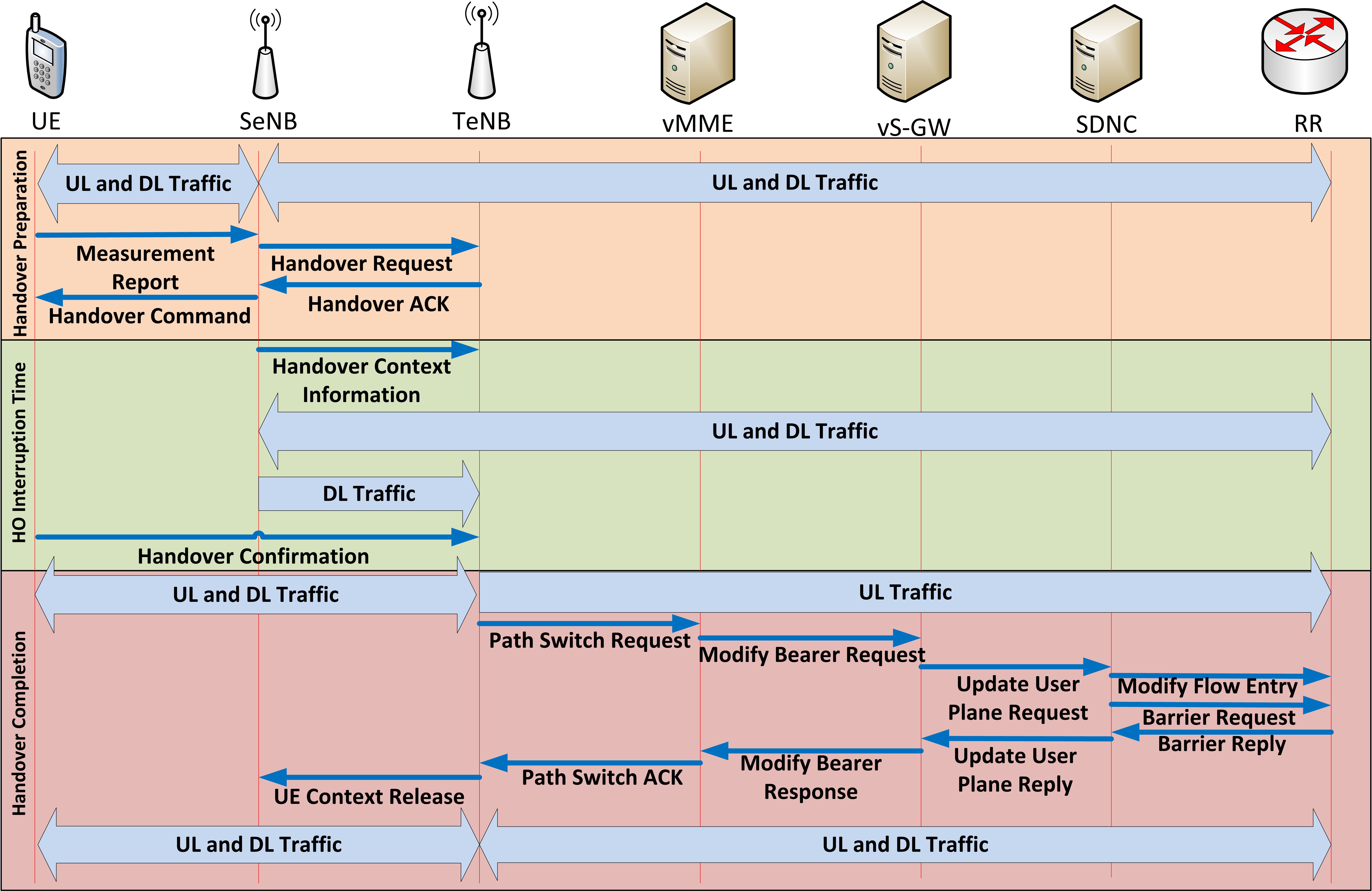}}
\caption{Proposed SDN based handover of \cite{prados2016handover}.}
\label{ho2procedure}
\end{figure}

As shown in the modified X2 call flow ( Fig.\ref{ho2procedure}) the radio link transfer involves the SDN controller transferring the path of the UE from the source eNB to the target eNB.

\subsubsection{Impact of SDN/NFV on the performance}

Using the proposed SDN based X2 handover, an Intra-Switch Handover and an Inter-Switch Handover are tested, through the NS-3 \cite{ns3} OpenFlow Module. The authors claim that the new system achieves a delay of 15.23ms from the handover decision to handover completion. This meets the Control Plane latency requirements for 5G.

\subsection{Centralized SDN with NFV for WiFi/LTE Multi RAT Handover}
\subsubsection{Architecture}
\cite{gharsallah2019sdn} presents a handover approach in Ultra Dense Networks (UDNs) consisting of both WiFi Access Points and LTE eNBs. Long handover delays, handover failures, and increased signaling overhead are some problems associated with mobility management in UDNs, leading to poor quality of service and dropped connections. The overall network visibility of the control plane is enhanced with information about network conditions such as bandwidth, delay, and packet loss and with mobile context information such as location, speed, and direction. The SDN based handover management mechanism reduces the number of handovers and provides seamless handover, by predicting the path of the users, in cases of UDNs with WLAN APs and LTE eNBs. The architecture used is similar to the central SDN based architectures seen in \cite{prados2016handover}, with SDN controller, handover management, and other components in the control plane, and the switches, APs, and eNBs in the data plane. The proposed software defined Handover Management Engine (SDHME) and NFV are run as applications, in the application plane, above the control plane.  

\begin{figure}[htbp]
\centerline{\includegraphics[width=\columnwidth]{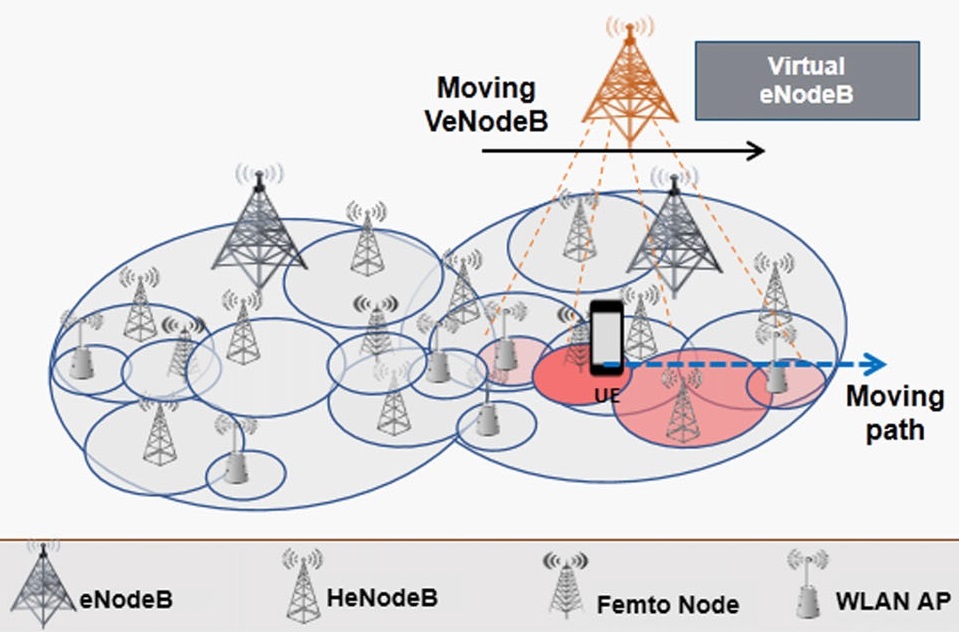}}
\caption{Proposed virtual moving eNB in \cite{gharsallah2019sdn}.}
\label{sdnnfvVeNbmoving}
\end{figure}

The authors propose to construct a virtual cell (V-cell) from a set of available base stations (Point of Attachments) of small and big cells, which are predicted to fall in the path of a user. Smaller cells are assigned for slow-moving traffic, and vice versa. So, for each user, a V-cell is constructed from the set. The V-Cell is a logical cell with a virtual eNodeB (VeNB) at its center, created and implemented through NFV. Thus, users having similar mobility profiles are assigned to the same VeNB. The VeNB predicts a given user's path and allocates relevant towers, initiating the handover in advance. So seamless handover is fulfilled, and the handover procedure is significantly optimized by reducing the handover delay and handover failure ratio. The Vcells, moving VeNBs for low-speed UEs are shown in Fig.\ref{sdnnfvVeNbmoving}.

\subsubsection{Handover Decision Process}

In the algorithm for software defined handover mechanism, received signal strength indication (RSSI), signal to interference, and noise ratio (SINR) received by the user due to a WiFi AP, and the SINR received in each subcarrier corresponding to each user are calculated. The calculation of RAT-specific SINR is a novelty seen in this algorithm. Additionally, while measuring the SINR in LTE, the mobile node is equipped with Global Positioning System (GPS) through which location(latitude, longitude), velocity, and direction are obtained. Using the positional information, the angle a mobile node's trajectory makes with the line joining the mobile node to the AP is calculated, thereby giving the direction of movement of the mobile node with respect to the AP and the positions of the APs and MNs.

Unlike the other papers seen in this survey, which focus on ranking the networks and selecting the best one, the objective is to determine the set of best APs that form the optimal anticipation pathway for the mobile node. The Software Defined Handover Management Engine achieves this using a cost matrix that denotes the quality of each network based on the RSS, Delay, Bandwidth, the Packet loss rate of each cell, and user mobility profiles such as speed and direction.

\subsubsection{Handover Radio Link Transfer Process}

This is not described.

\subsubsection{Impact of SDN/NFV on the performance}

The V-cell technique was compared with the existing LTE handover procedure through simulations in MATLAB. A mix of macro and small cells, with eNBs in macro cells and APs in small cells, is used. Tests measure the handover delay and handover failure ratio according to the increase in the number of mobile users. The results showed a 58\% decrease in handover delay compared with the standard mechanism and a 42\% decrease in handover failure ratio compared to the LTE protocol. To summarize, this paper is a novel approach that applies the benefits SDN offers, such as the ability to get a global view of the network, analyze the user characteristics and movement, and allocate network resources on demand using NFV, to tackle the handover problems in UDNs.

\subsection{Centralized SDN and Handover Decision Algorithms}

\subsubsection{Architecture}
Similar to predicting the path of UE, \cite{lee2017handover4g} proposes a linear programming based handover execution scheme that pre-allocates the cell the UE is going to move to, to reduce the handover delay. Although SDN is used to enable the seamless handover process, the type of SDN deployment is not specified. Since this paper talks about a single SDN controller, we consider this to be a centralized architecture, as this is the default architecture for SDN and the X2 handover architecture.

\subsubsection{Handover Decision Process}
In \cite{lee2017handover4g}, SDN is used to collect and store data from the network such as Reference Signal Received Power levels at the current cell and the neighboring cell; the number of nodes connected to a cell; the maximum number of nodes that can be connected to a cell; communication range of the cell; and the distances between two nodes (obtained from GPS information). Information about the UE such as the speed of UE, and the instantaneous angle UE's trajectory makes with respect to the cell (base station) is received from the UE itself. Using the collected data, the handover decision involves executing a linear programming based algorithm, where the SDN Controller must calculate 4 different values:  

\begin{itemize}
    \item Signal strength difference.
    \item Sojourn time (time the user will spend in the new cell after handover, which is determined through UE speed and range of the cell).
    \item Ratio of nodes connected to cell to the maximum number of nodes that can be connected.
    \item Moving direction of UE with respect to the base station.
\end{itemize}

Cell selection handover calculation value, P, is defined as the product of the above four calculated values, and the Linear Programming based problem solving technique is applied, to find the best cell to handover the UE. This is done by defining an objective function of P such that the selected cell maximizes the value of P. The cell with the highest value of P is selected as the cell to be pre-allocated for handover.

 \subsubsection{Handover Radio Link Transfer Process}

This is not described.

\subsubsection{Impact of SDN/NFV on the performance}
 The implementation details are not specific with respect to details on the platforms used to simulate the handover. This makes it hard to evaluate the proposed algorithm in any capacity. Thus, this work is considered only for ideological purposes rather than comparing it with any other algorithm that proposes a seamless handover or handover optimization based on predicting the UE's path, such as \cite{gharsallah2019sdn}.

\subsection{Semi-Centralized SDN  architecture for vertical handover}

\subsubsection{Architecture}
\cite{rizkallah2018sdnverticalho} uses a semi-centralized architecture that deploys an SDN controller in each of its sub-networks, which are made up of multiple macro and femtocells. While this proposal is supposedly meant for 5G HetNets, it considers the LTE architecture as the basis of the proposed semi-centralized SDN architecture. The access network component is again connected to OpenFlow   switches, which communicate with the SDN Controllers. However, the details about the modification to the core network or their deployment are not mentioned in \cite{rizkallah2018sdnverticalho}. Using this architecture, the authors propose a vertical handover scheme to improve the mobility management of users within 5G femto eNodeB Heterogeneous Networks (HetNets).

\subsubsection{Handover Decision Process}

Information used in the handover is the network context information, user context information, and service context information to maintain the quality of service. The authors calculate the quality of service using a metric that takes into account network and service parameters such as Received signal strength, bandwidth, and jitter. The SDN handles the control messages of the Mobility Management Entity (MME) and P-GW, reducing the exchange of messages and handshakes in the signaling flow. In the decision phase of the handover, a quality of service function is used to calculate the 'q' score for the network. This 'q' score is a weighted sum of different network parameters and service parameters such as Received Signal Strength RSS, bandwidth, and jitter, as shown in Equ.~(\ref{eq1}). $q$ is given by:

\small
\begin{eqnarray}
q = \hspace{-5mm} & w_{RSS}\ln(RSS)  +  w_d \ln(\frac{1}{d}) + w_{BW}\ln(BW)  \\ \nonumber
& + w_{type}\ln(type) + w_{BLER}\ln(\frac{1}{BLER}) + w_J\ln(\frac{1}{J}) 
\label{eq1}
\end{eqnarray}

\normalsize

\noindent where w\textsubscript{RSS}, w\textsubscript{d}, w\textsubscript{BW}, w\textsubscript{type}, w\textsubscript{BLER}, w\textsubscript{J} are weighting factors for received signal strength (RSS), delay (d), bandwidth, application type based priority (type), block error rate (BLER) and jitter (j), respectively. The sum of these weights is equal to 1. This approach of using multiple network specific criteria in the handover decision process is usually referred to as a Multi Criteria Decision Making Algorithm. Additionally, the UE speed is considered in the algorithm, before calculating the 'q' score, as it would be unnecessary to handover to a femtocell if a user is moving at high speed. \cite{rizkallah2018sdnverticalho}'s proposed vertical handover decision scheme is as follows: 

\begin{itemize}
    \item the UE measures the RSS of the source cell as well as the neighboring cells. If the RSS of the source cell is below a certain threshold and the user speed is below a certain speed threshold, then a handover is triggered.
    \item The first three neighboring cells having the highest RSS, which is greater than the source cell are chosen. Of these, the cell with the highest ‘q’ score is chosen as the target cell.
    \item If a UE has a speed lower than a pre-defined second speed threshold, or if the user is running a real-time service, the handover is initiated. 
\end{itemize}

\subsubsection{Handover Radio Link Transfer Process}

The radio link transfer process is handled by the SDN controller without the involvement of the source eNB and target eNB, similar to the modified X2 call flow seen in Fig.\ref{ho2procedure}.

\subsubsection{Impact of SDN/NFV on the performance}

The paper evaluates the proposed model by comparing the traditional RSS based handover between the macro and femtocells in a 5G HetNets scenario using the X2 based handover \cite{etsi3gppeutra2010} and the proposed SDN based Handover. The number of handovers, the signaling overhead, the handover delay, and the total throughput in the network is compared. The results show that the use of SDN reduces the signaling overhead and handover delay without adding complexity to existing network nodes. A decrease in the number of HOs and a network throughput increase is also observed. The usage of SDN reduces the number of handshakes and exchange messages in the handover, thus reducing signaling overhead. 

\subsection{Drawbacks and Shortcomings}

Each paper discussed in this section provides a novel contribution in terms of architecture, handover optimization approach, and handover decision schemes based on SDN. But these come with some limitations, discussed as follows.

\cite{prados2016handover} provides a detailed implementation of the SDN based X2 handover, significantly reducing the handover latency experienced in LTE networks. Apart from being an outdated/legacy LTE architecture, this work fails to address several other aspects of the handover process such as handover cell selection, decision criteria, etc. \cite{gharsallah2019sdn} provides a novel approach of creating virtual cells and using the SDN's global view and Virtualized network functions obtained through NFV in predicting the UE/mobile node's path through the network. This enables the controller to prepare for a handover ahead of time, enabling seamless handovers in an environment like the UDNs where traffic density is significantly high. While the idea is encouraging, it remains to be seen, how practical such an idea can be in actual deployments, considering the latency/processing delays introduced by computation of the parameters introduced by the algorithm such as the direction of mobile nodes with respect to AP's, and the time taken by the SDN to create the virtual networks and predict the paths of thousands of UE's independently.

Finally, while the results in \cite{rizkallah2018sdnverticalho} seem promising, the architecture of the network in the handover scheme is again, based on 4G LTE. A significant shortcoming is a lack of a detailed description of architectural changes associated with incorporating SDN with LTE, as seen in the other papers. The 5G NR architecture already comes with a segregated data and control plane, so it remains to be seen if the performance improvement seen in the results of \cite{rizkallah2018sdnverticalho} will still hold when the proposed vertical handover algorithm is implemented on top of a 5G NR based architecture like seen in \cite{abdulghaffar2021modeling},\cite{dhruvik2021design}.

The papers covered in this section are important innovations of the time before 5G was standardized, in how the handover can be optimized using SDN. Introducing a novel SDN based architecture for 5G, which was not released when these papers came out is the theme of these papers. The way SDN is incorporated is very simple, by implementing the core network's functionality in the SDN controller. This approach has some limitations. The way the core network's functionality has to be modified to accommodate the control and data plane segregation is not mentioned in any of these approaches. Furthermore, the specific details about the flow tables and modifications needed for interfacing the base stations with the OpenFlow enabled switches are not covered in these approaches.


\section{5G based SDN architectures}
\label{5gsection}

Moving toward later developments in the cellular standards, the 5G NR specifications released by 3GPP \cite{etsi5Gwithsdn} provide a separation between the control and data planes. This section discusses the changes made to the 5G NR architecture using a centralized SDN approach, and to improve handover using SDN. Each paper in this section provides a novel approach to using the SDN in optimizing the handover in different network types and scenarios such as mmWave and small cell 5G networks.

First, \cite{abdulghaffar2021modeling} provides an SDN enabled architecture based on the 5G NR standard, corresponding SDN modified versions of UE's initial attachment, and an Xn handover. Next, \cite{dhruvik2021design} proposes a new interface and a simple handover algorithm to address the problems linked to handover in 5G mmWave systems, such as frequent handovers and maintenance of the quality of service in the case of high mobility of UEs. 
\cite{ciciouglu2021multi} addresses small cell 5G networks through changes to the OpenFlow table in the OpenFlow protocol and a Selective Average Weighting handover algorithm.

\subsection{SDN with Xn based  Handover}

\cite{abdulghaffar2021modeling} serves as a good paper to understand the SDN operation in 5G networks. The authors introduce an SDN enabled architecture based on the 3GPP 5G NR Standard. Two control plane procedures, initial attachment, and handover are explained and the performance of the SDN based 5G architecture is evaluated based on end-to-end delay, throughput at the controller, and the resource utilization of the controller. 

\subsubsection{Architecture}

The architecture used in \cite{abdulghaffar2021modeling} is a very basic implementation of SDN using the 5G NR architecture. The 5G NR architecture separates the control plane and user plane, but there is no allocation for the SDN controller. In the 5G NR architecture, the UE and the gNB (base station abbreviation in 5G) are connected to the Access and Mobility Management Function (AMF) of the 5G core network. As the name suggests, the AMF handles the handovers of the UE and provides access control, registration, and mobility management. In the proposed SDN based architecture, to make the functionality of the AMF available through the SDN, the SDN Controller is also connected to the RAN or the gNB in addition to the UPF.

\begin{figure}[htbp]
\centerline{\includegraphics[width=\columnwidth]{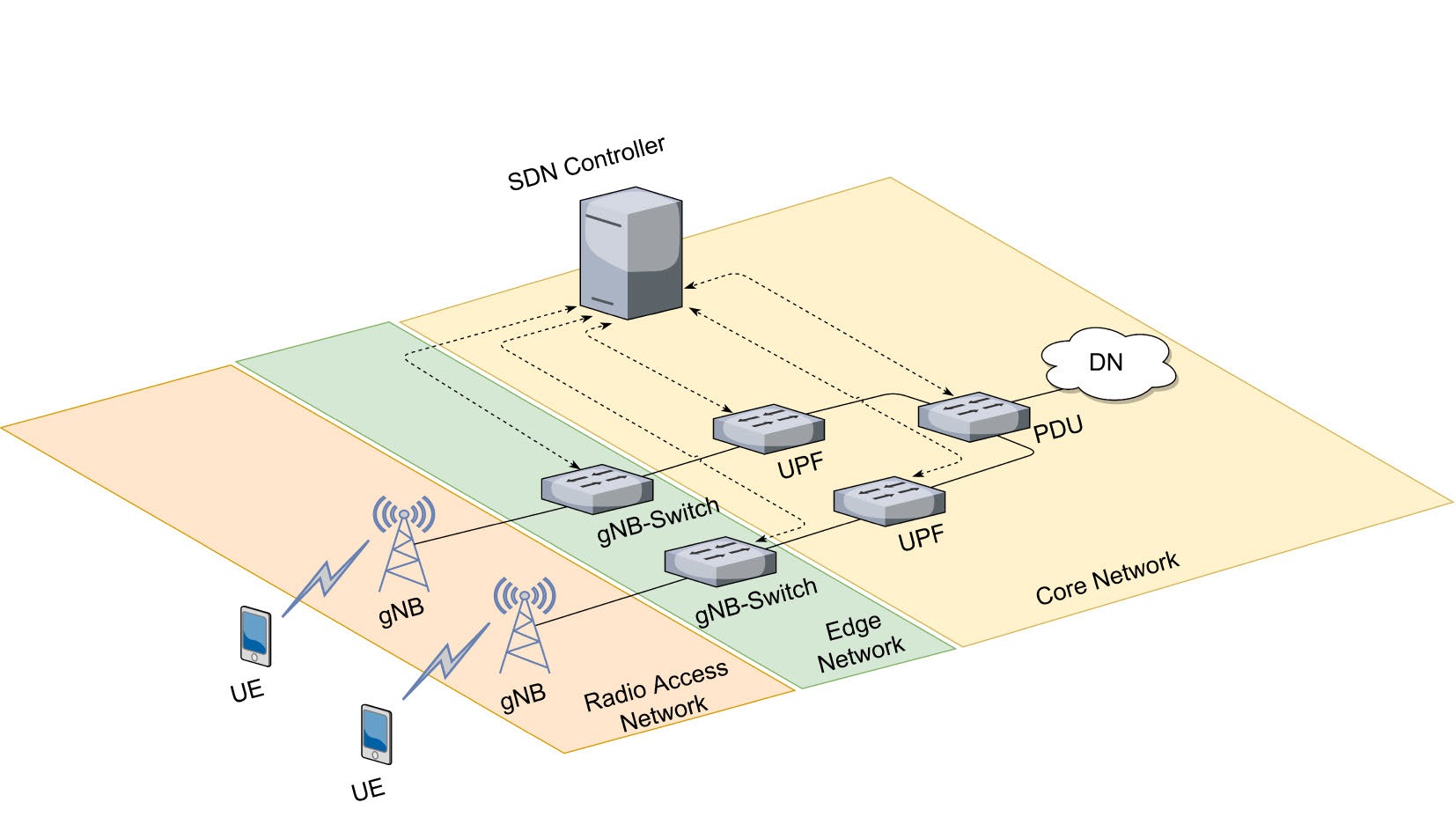}}
\caption{Implementation of \cite{abdulghaffar2021modeling}'s proposed architecture.}
\label{5Gpapimp}
\end{figure}

All network management functions of the 5G core network such as AMF, SMF, and AUSF \cite{etsi5GS2020} are moved to the SDN controller. All these control plane functions are implemented as applications and interact with the SDN controller via the northbound interface, shown in Fig.\ref{SDNarch}. The data plane of the 5G network, i.e. the UPF, is implemented with SDN switches. Therefore, UPF uses the OpenFlow protocol to communicate with the controller on the south interface, shown in Fig.\ref{SDNarch}. In order to make gNB compatible with the SDN switch, the authors assume that each gNB in the RAN is connected to an SDN enabled switch as shown in Fig. \ref{5Gpapimp}. The RAN of the 5G network is not modified, as the paper concentrates on the core network only.

\subsubsection{Handover Decision Process}

The handover process in \cite{abdulghaffar2021modeling} is very similar to the Xn based 5G handover process. The authors modify the Xn based 5G handover to include the SDN's functionality. Instead of the path switch requests going to the AMF, they are routed to the SDN Controller, which houses the AMF in the proposed architecture. Two handover procedures, with and without UPF re-allocation are explained, but the criteria for the handover decisions are not clear.

\subsubsection{Handover Radio Link Transfer Process}
This topic is not covered in this paper.
\subsubsection{Impact of SDN on the performance}

Mininet\cite{mininet} is used to validate the simulation along with an analytical model \cite{mahmood2015modelling} based on a system modeled using Jackson Network\cite{jackson1957networks}. A Ryu SDN controller\cite{ryu} is used as an SDN controller for the mininet emulation. In addition to the emulation, the authors build an analytical model based on Jackson's network\cite{jackson1957networks} to validate the findings obtained in mininet. 

The paper compares the performance of three procedures (registration, inter UPF handover, and Intra UPF handover) with the proposed model against the traditional 5G NR architecture. The results show a 53\% reduction of end-to-end delay for initial attachment, 24\% reduction in case of inter UPF handover (without UPF reallocation) handover, and 26\% reduction in case of intra UPF handover (with UPF reallocation)  when using the proposed architecture over traditional 5G networks. The performance can be attributed to the use of the SDN controller in the handover decision which reduced control messaging compared with the traditional architecture.

The centralized architecture that uses the 5G NR specifications is very common \cite{abdulghaffar2021modeling}, \cite{dhruvik2021design}, \cite{ciciouglu2021multi}. The cornerstone of such architectures is similar, with SDN orchestrating the network traffic, but the implementation of this architecture varies for each paper according to the objectives and problems they look to address. The main limitation \cite{abdulghaffar2021modeling} is the implementation of the modified Xn based handover and the absence of optimization in regards to handover decision criteria as well as no consideration of user context. From an implementation standpoint, the RAN is left unchanged, and only the core network functionality is simulated, so the results of a full-fledged SDN based Xn implementation still need to be evaluated. Additionally, the 5G based architecture and the 5G NR protocol stack were not considered in the evaluation of the proposed algorithm because such simulators are not available.

\subsection{SDN with 5G Millimeter Wave networks}

The introduction of the higher frequencies into the 5G spectrum i.e., the mmWave frequencies comes with a new set of challenges due to the nature of the radio waves. They are easily blocked by common building materials, and humans, causing the channel to rapidly change. It introduces smaller cells due to more rapid attenuation at higher frequencies. So, mobility management in such dynamic network situations becomes a challenge giving rise to repeated handover failures and ping pong effects. To address these issues, the authors of \cite{dhruvik2021design} propose an SDN based handover for the 5G mmWave, with a new handover interface called the HOinterface and bring SDN functionality to a mmWave simulation platform, called NS3-mmWave patch. 

\subsubsection{Architecture}
This paper uses an LTE based architecture, despite using mmWave frequencies. This is driven by the lack of an NS-3 5G protocol stack requiring the use of NS-3 mmWave patch based on the LTE protocol stack using mmWave frequencies. While the type of SDN architecture is not mentioned, the SDN is used as a single entity, in its default configuration. So this can be considered a centralized SDN architecture. The SDN controller is placed on top of the core network, giving the global view of the network. The proposed handover interface is a simple point-to-point interface, which connects the SDN controller with the P-GW. The purpose of the SDN is to collect information from UE and make an intelligent handover decision, as seen in other papers.

\begin{figure}[htbp]
\centerline{\includegraphics[width=\columnwidth]{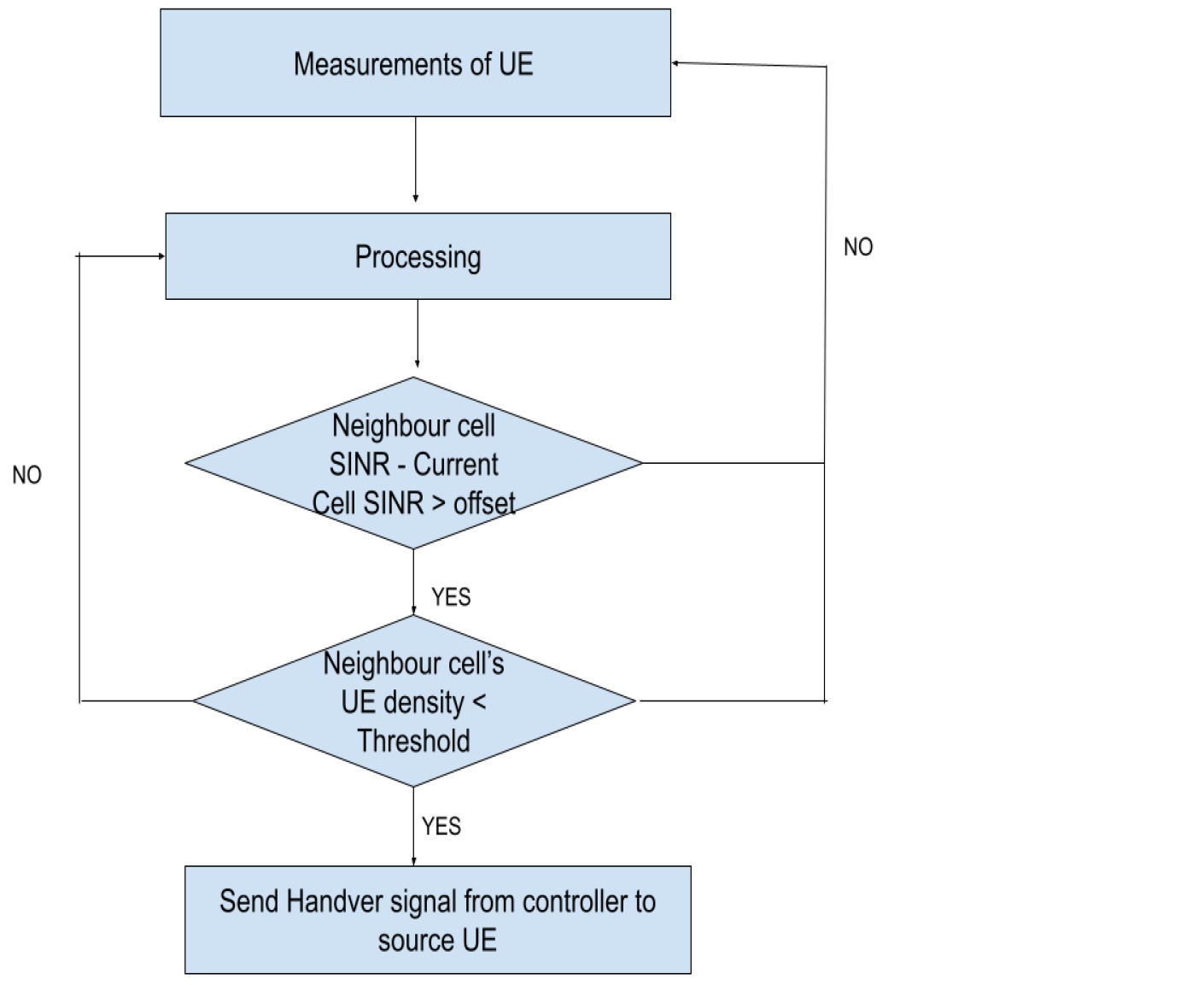}}
\caption{handover algorithm proposed in \cite{dhruvik2021design}.}
\label{5gmmwaveHO}
\end{figure}

\subsubsection{Handover Decision Process}

A measurement report from each UE, containing SINR with respect to eNBs, is sent to the controller periodically every 200ms. This value is set arbitrarily by the authors. The controller uses these reports from all the UEs in the network to make a handover decision. The proposed handover algorithm is shown in Fig.\ref{5gmmwaveHO}. For each UE, the controller checks if the neighboring cell's SINR value exceeds the current cell's SINR value by an offset. Handover is triggered only if the neighboring cell's user density is less than a predetermined threshold. Once a handover is performed, the target cell will become the current cell for the UE. This detail will get updated at all the places in the network. This entire process happens periodically in the SDN based handover algorithm.

\subsubsection{Handover Radio Link Transfer Process}
This topic is not covered in this paper.

\subsubsection{Impact of SDN on the performance}
To evaluate the proposed method, the authors use the NS-3 mmWave patch. They analyze the system performance based on Throughput, Number of Handovers, End-to-End Delay, and scalability. The results show the successful implementation of SDN based handover. unfortunately, these are somewhat general network performance tests such as throughput and delay increases with the number of nodes. The impact of SDN on handover performance has not been shown. It does verify that SDN can be implemented in a mmWave 5G network. Furthermore, the type of handover algorithm used is simple, similar to the one seen in \cite{rizkallah2018sdnverticalho}. While the authors claim the proposed algorithm is a decent one, several other papers such as \cite{ciciouglu2021multi} build on this idea. 

The choice of the simulation tool, NS3 mmWave patch is based on the LTE protocol stack, but not the 5G standards, as the paper suggests. Thus, this is also a case of a simulator based limitation, due to the non-availability of the tools required to simulate mmWave and 5G stack with SDN.

\subsection{SDN based Multi Criteria Handover in 5G small cells}
\begin{figure}[htbp]
\centerline{\includegraphics[width=\columnwidth]{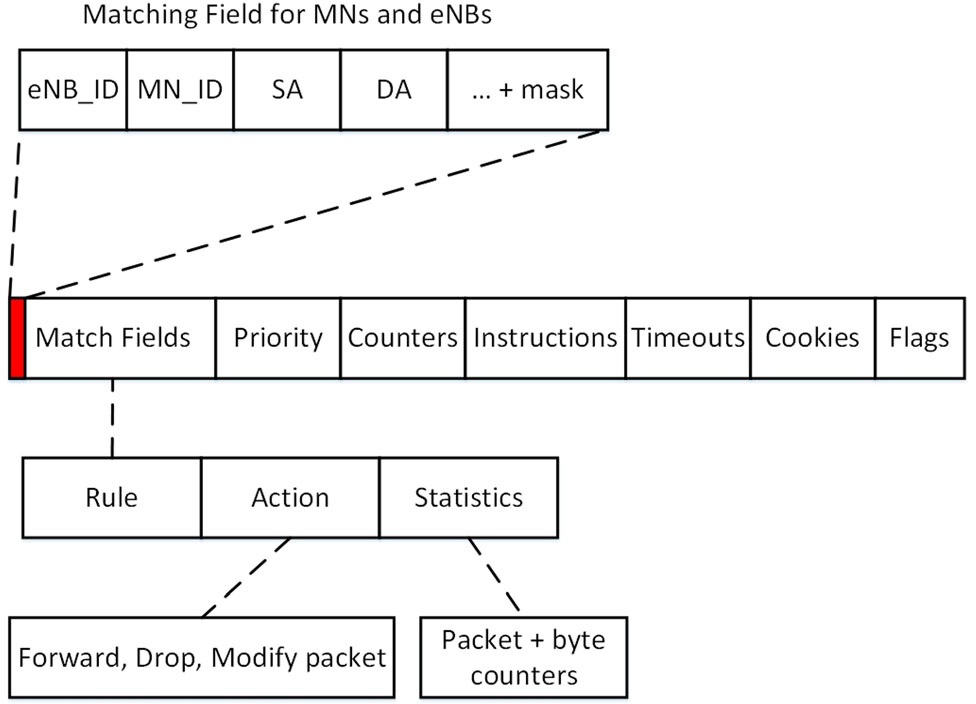}}
\caption{Modified OpenFlow message in \cite{ciciouglu2021multi}.}
\label{SAWOFmsg}
\end{figure}

Another emerging domain due to the inclusion of mmWave frequencies and the need to support higher data rates is that of small cells. In 5G, small cells are characterized by a small geographical area of coverage, and a massive number of such cells. Due to these characteristics, small cells are often faced with frequent handovers, interference, and delays. \cite{ciciouglu2021multi} proposes a centralized SDN based Architecture for small cell 5G networks, to address the handover management challenges. According to  \cite{ciciouglu2021multi}, a lack of well-defined boundaries and the use of sectoring, which causes overlaps between boundaries, causes nodes to be connected to multiple eNBs at the same time. Thus, it becomes complex to manage handovers and avoid multiple unnecessary handovers. 

\subsubsection{Architecture}

The authors explain that the control and management of small cells with low coverage becomes a complex problem with a distributed network approach. So, they use a centralized SDN architecture. They explain that centralized SDN approaches are more efficient, as the controller has information on the entire network and can make better decisions when it comes to handover management in small cells. The architecture used in \cite{ciciouglu2021multi} is similar to the architectures seen in \cite{abdulghaffar2021modeling} with SDN controller in the control plane and the 5G core functions running as applications on top of the controller. Communication between the controller and data plane devices takes place through OpenFlow protocol \cite{openflow}. 

unlike other papers, in \cite{ciciouglu2021multi}, the OpenFlow flow table has been modified to accommodate the proposed protocol. Parameters like IDs of mobile nodes and small cells have been added to the match field for the communication from control to the data plane, as shown in Fig.\ref{SAWOFmsg}. Decisions taken regarding handover management are transmitted to data plane devices by updating flow tables. 

\begin{figure}[htbp]
\centerline{\includegraphics[width=9cm]{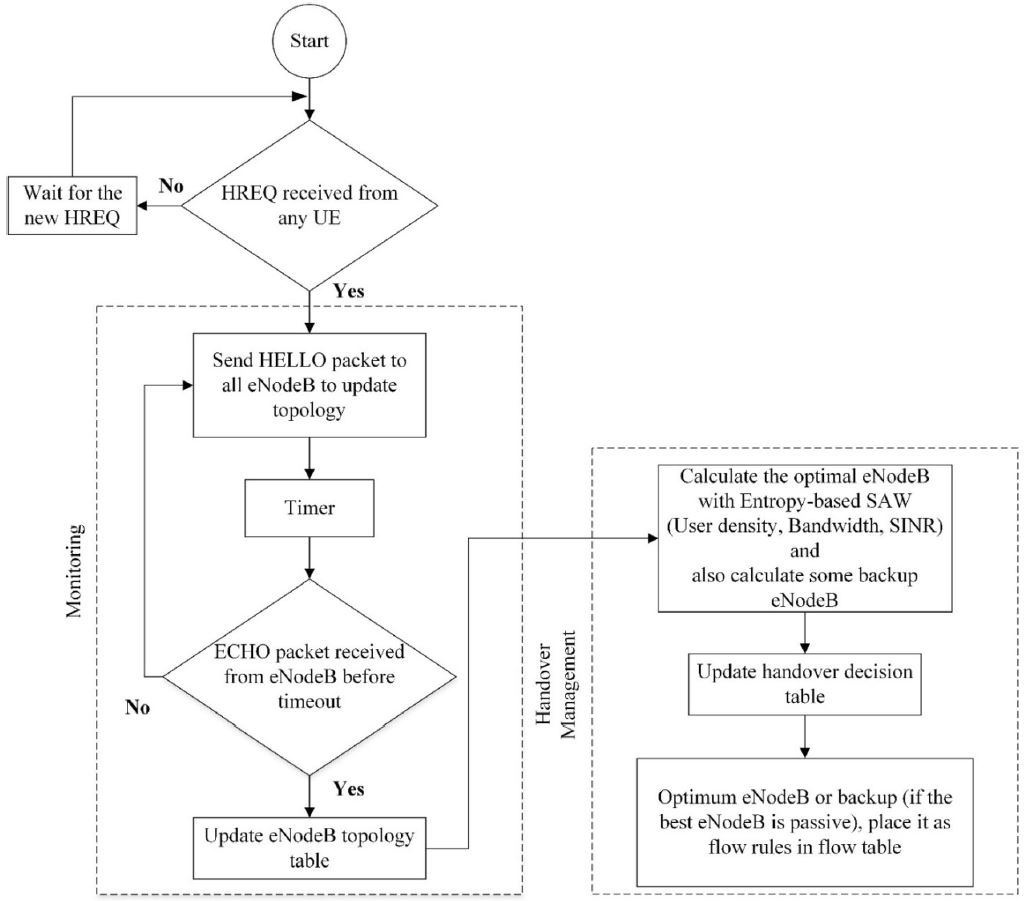}}
\caption{Proposed Handover algorithm of \cite{ciciouglu2021multi}.}
\label{SAWHOalgo}
\end{figure}

\subsubsection{Handover Decision Process}

When considering the handover for each UE, the controller has a list of all available eNBs for a given UE and the values of bandwidth, SINR, and user density at each corresponding eNB. These values are reported by all the UEs in the network to the controller periodically.
In the handover decision phase, the authors propose an entropy based Simple Additive Weighting(SAW) to determine when the handover is made. This is another type of Multi Criteria Decision Making scheme, similar to the one seen in \cite{rizkallah2018sdnverticalho}. According to the authors, in an entropy based method, the weight associated with each decision attribute is proportional to the information. If the difference between the attribute ranges is large, it contains more information and greater weight is assigned to this attribute. In this way, attributes that have been assigned the maximum weight are determined by an objective approach. No user priority is taken into account and the weight depends on the attribute range. 

Entropy based weights are obtained for each attribute, followed by the calculation of the SAW metric. The Simple Additive Weighting metric is obtained by taking a simple weighted sum of the obtained entropy based weights and the normalized attribute values of bandwidth, user density, and SINR of each available eNB. The available eNBs are ranked according to these to obtain a decision matrix, with the best eNB having the maximum Simple Additive Weighting score. A decision is made to switch this eNB with the highest Simple Additive Weighting score and is installed as a flow table rule in the switches by the controller. Fig.\ref{SAWHOalgo} shows the algorithm of decision making and procedure flow of the handover proposed in \cite{ciciouglu2021multi}. 

\subsubsection{Handover Radio Link Transfer Process}
This topic is not covered in this paper.

\subsubsection{Impact of SDN on the performance}

OPNET(Riverbed Modeler)\cite{opnet_riverbed} is used to test the proposed handover decision scheme and test it with respect to the LTE Handover algorithms. handover delay, handover blocking probability, failure ratio, and throughput is compared with each algorithm while increasing the number of mobile nodes. The results show that the proposed Simple Additive Weighting based handover outperforms the LTE conventional handover by 15\% in handover delay, 48\% in blocking probability, and 22\% in throughput respectively. The improvement in the performance over LTE can be attributed to the entropy based Simple Additive Weighting and the SDN controller that collects the information and computes the SAW decision metric in real time.

This paper is a great addition to the research on the ways a handover decision can be made using SDN. This paper does not provide priority or consideration to application types. It considers the lack of information about a network attribute, as less information, and objectively weights the parameters. Additionally, unlike the Multi Criteria Decision Making scheme shown in \cite{rizkallah2018sdnverticalho}, where weights are calculated by the user, this method of weight selection is reflective of the network characteristics and can be dynamically adjusted automatically without user interference. While simulation results seem promising, exact information on which LTE conventional procedure was used to run comparisons isn't provided, thereby making these observations less specific and less comprehensive.

\subsection{Observations and Limitations}

The papers in this section introduce SDN into the 5G NR architecture to address problems in specific 5G network scenarios. \cite{abdulghaffar2021modeling} provides a basic SDN based modification to the Xn handover, which serves as a good starting point for handover optimization as the incorporation of SDN in itself improves the handover according to the experimental results. A significant limiting factor is the non availability of a common platform for simulating SDN and 5G NR protocols together, as the authors of \cite{abdulghaffar2021modeling} only consider the core network, leaving the access network unchanged. 
\cite{dhruvik2021design} proposes an inconclusive handover algorithm for the case of mmWave, but highlights the lack of availability of mmWave simulators with 5G protocol stack. Despite using the LTE architecture in their proposed architecture, the use of the handover interface to bridge SDN with the LTE EPC to simulate a handover in mmWave network seems to be effective. Finally, \cite{ciciouglu2021multi} proposes an 5G based algorithm with SDN integration, similar to \cite{abdulghaffar2021modeling} and addresses the concern of simple handover that we underline in the algorithm used in \cite{abdulghaffar2021modeling}. Specifically, \cite{ciciouglu2021multi} introduces the Selective Average Weiging method to dynamically select weights for network parameters without user interference. 

The papers in this section either cover the SDN based architecture with a simple handover algorithm \cite{dhruvik2021design} or a comprehensive handover decision that is based on a generic architecture\cite{ciciouglu2021multi}. There is no single paper with a comprehensive SDN architecture description and how the architecture coupled with a comprehensive algorithm can improve the handover performance compared to the existing 5G standard. A centralized SDN architecture is prone to a single point of failure, and the single controller acts as a bottleneck preventing network scaling. Next, there is no information as to how network functions are implemented on top of the SDN controller, without the use of NFV. This is a significant limitation for reproducing or extending the research presented. The 5G NR standard accommodates the SDN by separating the control and data planes inherently. SDN is used along with NFV to allocate resources dynamically for network slicing in the 5G NR standards. This aspect is not explored. Finally, a common shortcoming of these works is the lack of a cohesive simulation tool to implement the protocols on the 5G NR protocol stack and SDN.

\section{SDN  for Multiple Radio Access Technology (RAT) Mobility}
\label{multiratsection}

Multi RAT networks refer to the co-existence and or co-location of multiple types of Radio Access Technologies. These can be under the same set of standards. 5G can consist of sub 6 GHz and mmWave, in addition to providing support to legacy RATs, LTE, UMTS, and GSM. In addition 5G may coexist among other non-cellular wireless RATs, such as WiFi, LoRaWAN\cite{lorawan}, IoT networks and Bluetooth.

The central promise of SDN based Multi RAT networks is the ability to interface the multiple RATs through an SDN controller and use the other RATs to improve the quality of service or experience. In the previous sections, the architectures of established standards, such as 5G and LTE, are modified to incorporate SDN. However, the incorporation of SDN into Multi RAT networks offers a different challenge: the non-compatibility of the standards of each RAT. Hence, apart from the handover optimization, details about how the SDN enables the multi RAT architecture are also covered.

\cite{taksande2020open5g}, proposes modifications to the 5G NR architecture, to truly separate the control and user planes in the Next Generation RAN (NGRAN). It also integrates the Wireless LAN (WLAN) into the NGRAN making it a Multi RAT implementation. Next, \cite{wang2017sdn} approaches the multi RAT paradigm through a novel Virtual RAT architecture. This is made possible using software based interface sets controlled by an SDN controller to deal with heterogeneous RATs and additionally, to optimize radio resource allocation among the RATs. Staying along the same lines of handover optimization using SDN in multi RAT networks, we cover papers \cite{Alfoudi2019seamless}, \cite{alotaibi2018hierarchical},\cite{alotaibi2021linking} that use hierarchical SDN architectures to address different issues in multi RAT networks. In \cite{Alfoudi2019seamless}, distributed controllers in a hierarchical architecture are managed using the Distributed Hash Table (DHT).A seamless mobility management method is proposed that uses both homogeneous and heterogeneous handover among the available RATs, WiFi, LTE, and 5G. \cite{alotaibi2018hierarchical} proposes to use a vertical Hierarchical based SDN architecture in conjunction with the Mobile IP based handover mechanism. \cite{alotaibi2021linking} uses a similar hierarchical architecture, and proposes a heuristic approach to balance the load among the controllers, to reduce the controller’s response time. Each of the three papers uses a similar hierarchical architecture but uses the advantages offered by this architecture to solve different aspects of the handover optimization process. Finally, \cite{liyanage2017sdnoffloading} provides an operator-assisted approach to offload LTE data onto WiFi using SDN, while \cite{wang2016convergence} provides a convergence architecture for LTE and WiFi networks, to achieve a seamless vertical handover.

\subsection{SDN based Multi RAT RAN Architecture}

While there is a separation of the control/data plane in the current 3GPP 5G RAN architecture, a standard protocol for the management of data-plane functions of different RATs doesn't exist. In 5G, UEs can connect to WiFi, i.e. the WLAN through a separate Non-3GPP Interworking Function (N3IWF). There is no direct integration of the WLAN with the NGRAN. Additionally, there is no logical separation between the network control and UE control functionalities in the 3GPP NGRAN, like the separate control in 5GC through the Access and Mobility management Function (AMF) and Session Management Function (SMF). UE control and authentication in the 5GC is handled by the AMF while the network control and control of the user plane function (UPF) are handled by the SMF. To solve these issues, the authors of \cite{taksande2020open5g} propose an SDN based Multi RAT RAN architecture, that segregates the control plane of the RAN nodes from their data planes, and proposes a novel protocol called Open5G, that serves as the interface between the proposed SDN based Multi RAT RAN and the data-plane nodes. 

\subsubsection{Architecture}
\begin{figure}[htbp]
\centerline{\includegraphics[width=\columnwidth]{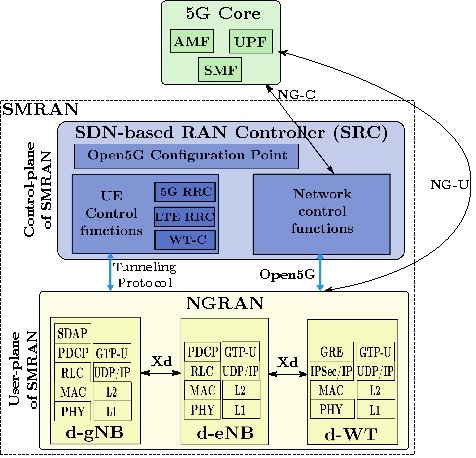}}
\caption{SMRAN: SDN based Multi RAT RAN architecture for 5G as seen in \cite{taksande2020open5g}.}
\label{open5Garch}
\end{figure}

As shown in Fig. \ref{open5Garch}, the proposed architecture has a logically centralized SDN based RAN Controller and an NGRAN with the various data-plane entities of different RATS. The following additions/modifications to the existing 5G NR protocols are made:
\begin{itemize}
    \item The network control functions of SDN based RAN Controller communicate with the 5G core using the Next Generation control-plane (NG-C) interface, which uses a protocol stack similar to the ones found in the AMF.
    \item The network control messages to configure the RAN nodes are sent using Open5G. 
    \item The UE control functions block provides the UE-specific control functionality of the RATs as shown in Fig.\ref{open5Garch}
    \item A tunneling protocol is used to transport the control messages from the SRC to the RAN data-plane nodes. From there, the corresponding signaling radio bearers (SRBs) are used to send them over the radio interface.
\end{itemize}

 \begin{figure}[htbp]
\centerline{\includegraphics[width=\columnwidth]{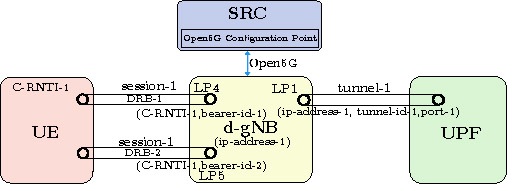}}
\caption{Open5G ports configuration for PDU session, as seen in \cite{taksande2020open5g}.}
\label{open5gPDUportsconfig}
\end{figure}

\begin{figure}[htbp]
\centerline{\includegraphics[width=7cm]{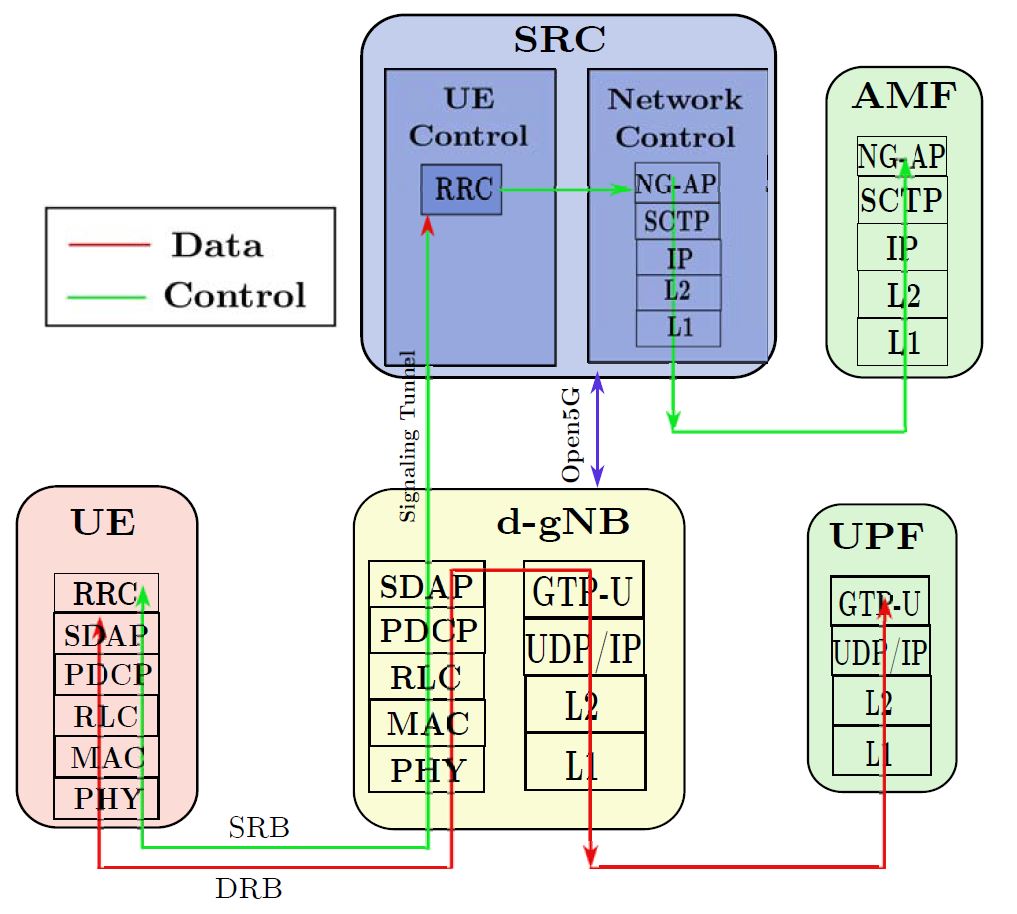}}
\caption{Data and Control Signal path in the proposed SMRAN and Open5G of \cite{taksande2020open5g}.}
\label{open5Gsignalpath}
\end{figure}

The architecture only provides a unified RAN with allocations to other RATs. Open5G is a modified version of the OF\cite{ovsv1.5.1} and OF-Config protocols\cite{openflow1.2}, with concepts such as Flow tables, physical and logical ports mapped for wireless networks that are based on the proposed SMRAN architecture. the Open5G protocol is responsible for the establishment of PDUs between the UEs and data plane entities of the RATs, and also for the control signaling path between the UE and the SRC. Fig. \ref{open5gPDUportsconfig} shows the establishment of a PDU session with a 5G NR UE and its port configuration. For the PDU session shown, the direct connections between the UE and d-gNB make up a physical port and the link between each bearer and UE make up the logical port. Data Radio Bearers and Signal Radio Bearers are generated according to requirements for the data (PDU) session and signaling sessions respectively and are established through the logical ports of d-gNB and UE's. On the other side, The GTP tunnels (tunnel-1 shown in Fig.\ref{open5gPDUportsconfig}), in case of data, corresponding to individual session paths for a UE towards the 5GC are used to identify the logical ports on the UPF side of d-gNB. Fig.\ref{open5Gsignalpath} shows the data and control signal path using Open5G. For the signaling messages, the link from d-gNB to SRC is treated as another logical port that forms a tunnel to SRC. This way, Open5G port configuration is carried out between nodes belonging to multiple RATs. The authors also explain the algorithms for UE initial access using SMRAN and Open5G, with detailed explanations of flow rules, logical and physical port configurations, etc.

\subsubsection{Handover Decision Process}
The authors of \cite{taksande2020open5g} explain the architectural changes and the new Open5G protocol. They claim that the Open5G interface allows the connection between UE and gNB to be configured as a logical link, this can be extended to connections between nodes in WiFi, i.e. node to d-WT using the same Open5G protocol. No details are provided about the establishment of connections to the other mentioned RATs in the NGRAN such as WiFi. The paper lacks details on how WiFi sessions are established or how a node in the 5G network can move into an LTE network. 

\subsubsection{Handover Radio Link Transfer Process}

Information on the transfer of the radio link between the different RATs supported in the proposed SMRAN is not provided.

\subsubsection{Impact of SDN on performance}

This paper provides a conceptual framework, without any evaluations or experimental analysis. Using SDN, the authors create a new architecture and the Open5G interface, which provides a unified interface for the different RATs. The limitation is the lack of evaluation or proof of concept of the proposed multi RAT architecture and a lack of information on handover between these multiple supported RATs.

\subsection{SDN based Virtual RAT for Multi RAT networks}

In \cite{wang2017sdn}, a new approach to achieving multi RAT networks is proposed through an SDN based virtual RAT architecture. The traditional methods of realizing multi RAT networks are standalone and integrated RATs, which have significant limitations \cite{multiRATlimitations}. In standalone RATs, different services are provided through various RATs that work alone. To switch between RATs, manual operation is needed. In integrated RATs, switching from one RAT to another requires coordinated and triggered automatic switching. Although some level of coordination and automatic switching is possible, it comes with high cost and  complexity because the RATs are not combined at the hardware level. To overcome these, \cite{wang2017sdn} proposes a Virtual RATs design using interface sets controlled by an SDN Controller to deal with heterogeneous RATs and to optimize radio resource allocation. Furthermore, they present self-healing mechanisms targeted at different failure cases in backhaul connections, which is out of the scope of this survey.

\subsubsection{Architecture}

The architecture of the proposed virtual RAT network is a mix of traditional macro cells and various types of small cells. Through a centralized SDN, the SDN Controller programs the interfaces and optimizes radio resource allocation. The idea is to program the behavior of communication elements like APs and switches through application-specific programming. Three programmable interface sets S1,S2,S3 are defined. S1 is between a small cell and an AP in traditional cellular networks. S2 is among small cells, while S3 is between relays and small cells. In a specific scenario, one interface will be selected and then programmed by the SDN Controller to perform the corresponding operations. Modifications will be updated in both the flow tables of the switch and the data center, where these interfaces are stored. A common programmable protocol stack is introduced, revised from the traditional one used in the 3GPP network. It is converted to a software package corresponding to a specific RAT. Software packages, supporting different kinds of services for various RATs, run on the same hardware platform to simplify the integration between different RATs.

\subsubsection{Handover Decision Process}

Handover between the virtual RATs occurs with the SDN Controller predicting the movement of the user and preparing the resources in advance. The authors introduce user-specific attributes called user contexts, which contain identity, signature, and user mobility information. With the overall view of the SDN Controller, the user contexts can be collected to predict the movement path and enable the handover to be done in advance. In a case where the predicted path is not followed by a user, handover using normal means is performed. 

\subsubsection{Handover Radio Link Transfer Process}
This aspect of the handover is not described.

\subsubsection{Impact of SDN on performance}

To evaluate the proposed mechanism of Virtual RAT, \cite{wang2017sdn} uses MATLAB simulations of a macro cell environment with multiple small cells, and users moving across these cells. The handover latency of proposed and traditional algorithms is compared with respect to the network load, defined as the ratio of the arrival rate and the SDN Controller's processing rate. The authors claim that while traditional authentication handover latency grows faster in latency, their proposed method maintains a latency of about 1 ms, thereby claiming their method outperforms the traditional methods.

This work provides a new approach toward multi RAT implementations with the ability to program the RAT as needed, to suit the type of connection that is required in a multi RAT scenario. There is a lack of unified access provided to the WLAN in the NGRAN. \cite{wang2017sdn}'s virtual RAT concept seems a valid alliterative to integrated RATs. However, integration of the proposed virtual RAT with the NGRAN is not specified, nor are the technical details involved in predicting the path of the users.

\subsection{Hierarchical SDN architecture for multi RAT networks}

As described in Section~\ref{background}, hierarchical architectures consist of different levels of controllers and are advantageous compared to centralized or distributed architectures. It allows the root controller to have a global view while allowing distributed controllers of subsequent levels to handle increasing traffic growth, hence improving scalability. Intuitively, the controllers on the lowest level can manage individual RATs, and the central controller on the topmost level manages these domain-specific controllers. Together, the network has the global view of a wide geographical location and all the involved RATs. Various papers approach controller management and use the proposed hierarchical approach to optimize various aspects of the multi RAT handover. \cite{Alfoudi2019seamless} proposes a hierarchical SDN based seamless mobility solution for multi RAT networks.  \cite{Alfoudi2019seamless} uses a Distributed Hash Table to manage communication between the hierarchical controllers. The authors also propose a homogeneous and heterogeneous handover in the multi RAT HetNet. 

\subsubsection{Architecture}
\begin{figure}[htbp]
\centerline{\includegraphics[width=\columnwidth]{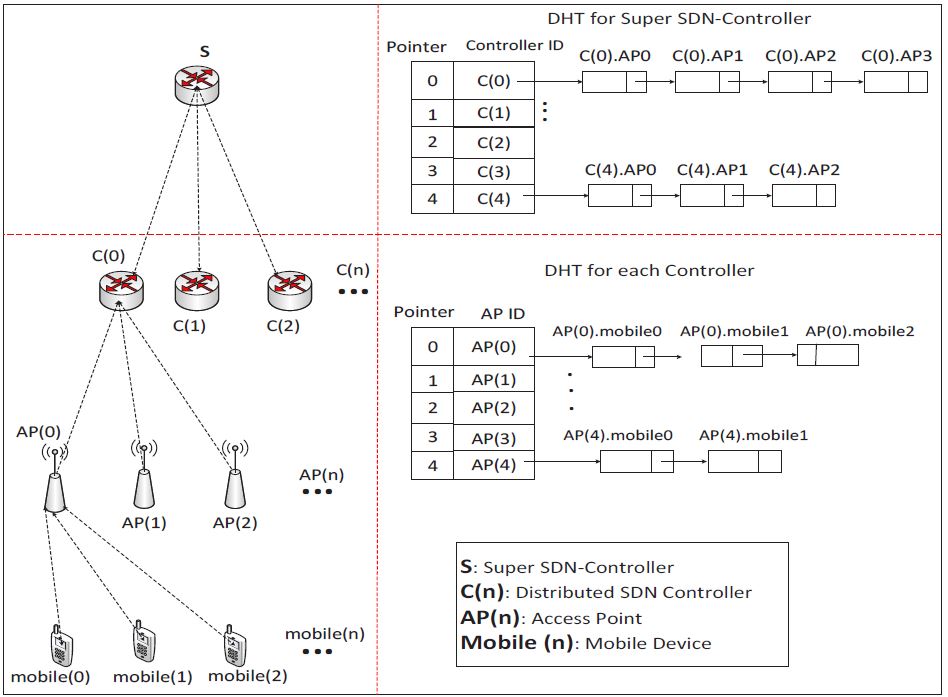}}
\caption{Distributed Hash Tables, as seen in \cite{Alfoudi2019seamless}.}
\label{DHTtables}
\end{figure}

\cite{Alfoudi2019seamless} uses a Hierarchical architecture as shown in Fig.\ref{DHTtables}. The super SDN controller at the top is in charge of controlling the distributed controllers, below it. Each distributed controller is connected to an independent domain of the network. Each SDN controller has a Distributed Hash Table table where all networking information is stored. The SDN end nodes in the hierarchy update their flow tables periodically or dynamically depending on network conditions and send the updated flow table to the main controller, which calculates required routing policies and distributes the decision of interconnectivity between them. The Distributed Hash Table for the super controller consists of the underlying SDN controllers, each AP associated with this controller, while the Distributed Hash Table for SDN controller has the APs associated and mobile users. These tables, as shown in Fig.\ref{DHTtables}, are updated with each handover request.

\subsubsection{Handover Decision Process}
The authors of \cite{Alfoudi2019seamless} propose two types of handover, homogeneous and heterogeneous.  Handover in both cases is initiated by the UE when it is in an overlapping area of two cells. The UE sends a message which has a list of nearby AP RSS and other parameters specified by location, time, and current service. The list of available APs is sorted in descending order of RSS. In the case of heterogeneous handover, the current AP forwards the handover request message of the user to the SDN controller which sends it to a "multi interface part" in the gateway. The details about the gateway are not provided. On receiving an acknowledgment, the current AP starts the handover steps.

\subsubsection{Handover Radio Link Transfer Process}
This topic is not covered in this work.

\subsubsection{Impact of SDN on performance}

The authors use OMNET++\cite{omnet} to compare handover latency and signaling message overhead of the proposed mechanism with two non multiRAT hierarchical Mobile IP mobility management approaches in literature \cite{seamlesscomp1}, \cite{seamlesscomp2}. Each mobility management approach is used in a LTE-WiFi handover and WiFI-WiFi handover. The results show that, for the LTE-WiFi handover, the proposed method has less handover latency than the other two mobility management schemes, as the binding updates related to the mobility of UEs are done only in the LTE-WiFi controller. Additionally, the signaling overhead through the proposed method is also the least, due to the global address of the UE remaining unchanged, which reduces the control signaling exchange between the UE and the Core network.

This work is useful in outlining the uses of a hierarchical architecture and the simplified mobility management this architecture brings with it. The Distributed Hash Table based mobility management is novel and not seen in other papers covered in this survey. But there are significant limitations to this paper. Information on the "multi interface part" which is responsible for handling the heterogeneous handover between different RATs is not mentioned. While there is an implementation of an actual LTE-WiFi multi RAT HetNet using OMNET++, the comparison with  non multi RAT mobility management schemes is not justified. Finally, this is a paper published before 5G standards were not released, so is based on LTE standards. It can be extended to match the existing 5G NR standards. 

\subsection{Hierarchical SDN with Mobile IP}
\begin{figure}[htbp]
\centerline{\includegraphics[width=\columnwidth]{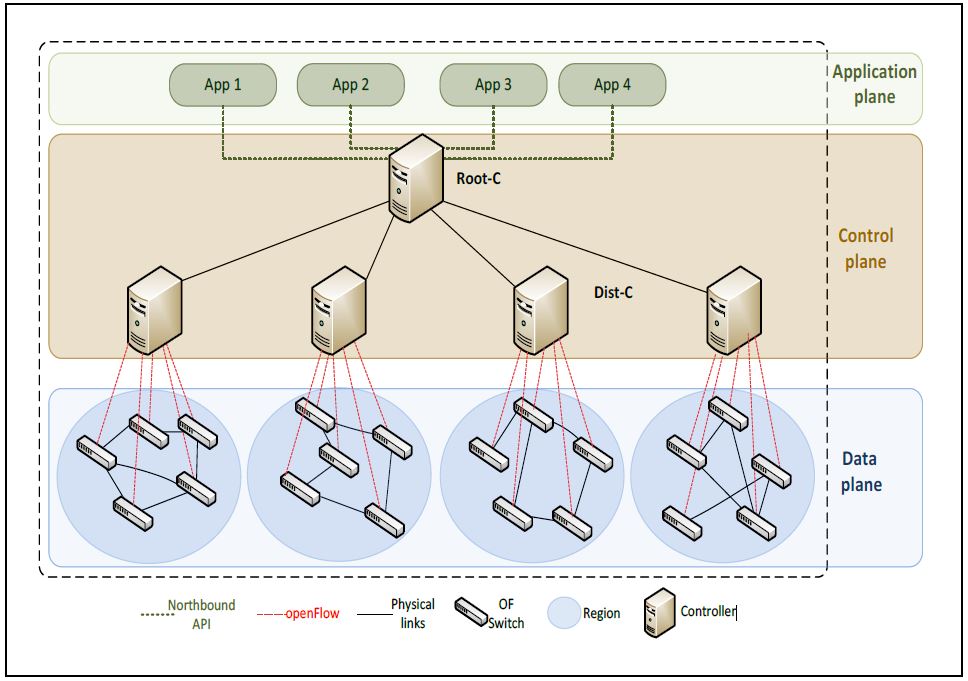}}
\caption{SDN hierarchical architecture in \cite{alotaibi2018hierarchical}.}
\label{hierarchicalMIParch}
\end{figure}

\begin{figure*}[t!]
\centerline{\includegraphics[width=\textwidth]{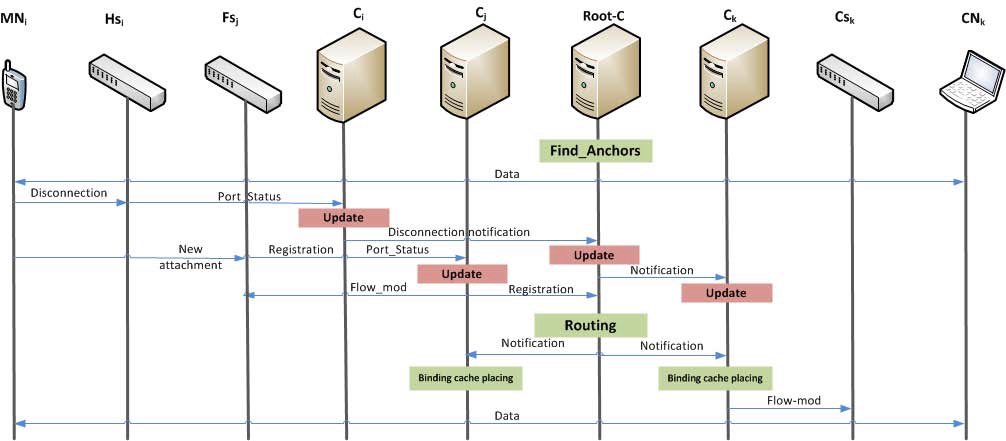}}
\caption{Handover protocol after a session initiated, in \cite{alotaibi2018hierarchical}.}
\label{hierarchicalMIPHOscheme}
\end{figure*}

As shown in Fig.\ref{hierarchicalMIPHOscheme}, \cite{alotaibi2018hierarchical} also bases its handover management in a multi RAT network on a Hierarchical SDN architecture. To overcome challenges such as scalability, security, and availability, the authors propose dividing the network into multiple connected regions instead of having one network controlled by a single controller. Different regions represent different service providers of one technology, or different technologies, making this a multi RAT implementation. To scale out the architecture, \cite{alotaibi2018hierarchical} follows a vertical approach, adopting two levels of controllers. Additionally, the authors propose a modified version of Mobile IP, a well-established protocol.

\subsubsection{Architecture}
\label{multirathierarchicalpaper1}

The authors of \cite{alotaibi2018hierarchical} propose Hierarchical SDN controllers of two levels: Master controller root controller, and a set of distributed controllers. The root controller manages a set of distributed controllers and each distributed controller manages a set of forwarding devices and switches in a particular region. 
\begin{figure}[htbp]
\centerline{\includegraphics[width=\columnwidth]{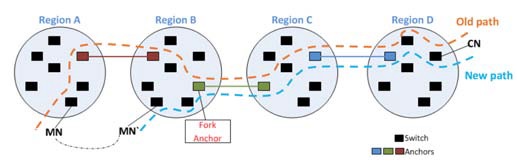}}
\caption{Fixed Anchors between Regions, and the difference between new and old path as explained in \cite{alotaibi2018hierarchical}.}
\label{hierarchicalMIPanchorrouting}
\end{figure}

\subsubsection{Handover Decision Process}

The authors propose the need to merge SDN and Mobile IP, in order to leverage the advantages of both. The authors consider the case where a mobile node moves from one region to another and communicates with a stationary Corresponding Node that belongs to a third region. When a node leaves a switch (Home Switch(HS)), the home switch reports the disconnection to its distributed controller through the 'port status' message. On the other hand, an attachment to a new switch, called a Foreign switch is notified to its distributed controller and then the root controller. The detailed protocol of such a session is shown in Fig.\ref{hierarchicalMIPHOscheme}. In the handover algorithm, authors define 'Anchor switches' among switches to reduce the number of updates that need to be done on switches during a handover. These anchors are determined using a distance function, defined by the authors, which considers both geographical distance and type of protocol in case of heterogeneous traffic. By having predetermined anchor switches between regions, they affix part of the path to get a prefix that does not need to be updated after a handover takes place. So a part of the old path that the mobile node was following will remain unchanged because only the part of the path where the mobile node has moved into different regions, i.e. a new Anchor will be the update needed to send to the switches.

\subsubsection{Handover Radio Link Transfer Process}
This is not covered in this paper.

\subsubsection{Impact of SDN and Mobile IP}

To evaluate the proposed handover mechanism, Mininet is used. The simulation topology consists of a root controller and two distributed controllers within the hierarchical architecture, and two mobile nodes, moving from one distributed controller's domain to another. The proposed mechanism is compared with a distributed architecture, with TCP socket as the messaging system to get messages moving between POX controllers. The distributed architecture's mobility management protocol details are not mentioned. A mobility scenario with the host moving between two switches, while monitoring the throughput and ping to monitor packet loss is implemented. Results show that both settings experience high jitter, but the hierarchical setting restores its steady rate of flowing packets faster. Additionally, the hierarchical architecture experienced 3\% packet loss, whereas the distributed architecture showed a slightly higher percentage, almost 3.4\%.  

The proposed Hierarchical SDN approach and anchor points within the network add a new implementation of the traditional anchor points idea. The idea of estimating the path of a UE is similar to the approach proposed in \cite{wang2017sdn}. The difference between the two is the type of architecture used, with \cite{wang2017sdn} using an architecture that can be virtualized, and a set of programmable interfaces, and \cite{alotaibi2018hierarchical} providing a more comprehensive way of predicting the path using the hierarchical architecture.

\subsection{Load balancing in Hierarchical SDN architectures to improve handover}

Extending the  work in \cite{alotaibi2018hierarchical}, \cite{alotaibi2021linking}, addresses an important aspect of handover processing, the processing delay of signaling messages. The authors explain that the load at the controller plays a crucial role in processing delays. So the paper seeks to minimize the maximum response time of a controller and attain timely delivery of large sizes of control messages by exploiting user preferences and complementary resources available at other controllers, by designing a load-balancing framework. This is based on a hybrid SDN architecture with multiple SDN controllers in the control plane, each controller connected to individual networks which may be of different radio technologies, making this a multi RAT implementation. 

\subsubsection{Architecture}

The architecture used is the same as the hierarchical architecture of \cite{alotaibi2021linking}, described in Section \ref{multirathierarchicalpaper1} with two levels of controllers, the root and distributed controllers.

\subsubsection{Handover Decision Process}

The handover problem is simplified to the processing delay at the controller due to a large number of signaling messages. The authors define an objective function, using detailed system modeling, to minimize the maximum response time and arrive at a set of constraints for each controller. The set of constraints defined are as follows: No controller is overloaded, i.e. load is always below a predefined threshold, each switch can be managed only by a single controller, each user has to be connected to only one switch at a given time, each switch has access to a limited number of channels for users to access.

To arrive at these constraints, a load-balancing mechanism and framework is proposed, which addresses the following:
\begin{itemize}

    \item Relieve the overloaded controller by exploiting context-aware vertical mobility when applicable, between different kinds of network technologies.
    \item Reduce status synchronization overhead.
    \item Optimize network selection for handed-over users, using Analytic Hierarchy Process from the first objective.
\end{itemize}

\begin{figure}[htbp]
\centerline{\includegraphics[width=9cm]{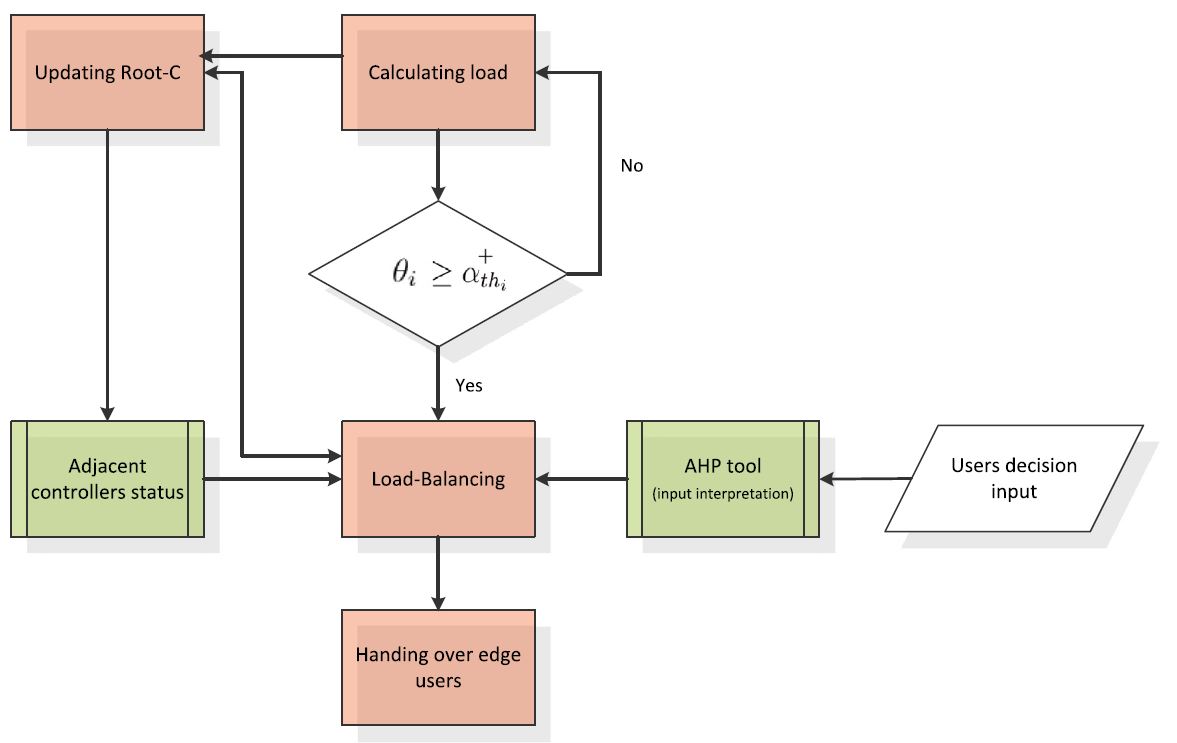}}
\caption{Load Balancing Flowchart, as proposed by \cite{alotaibi2021linking}.}
\label{loadbalancingalgo}
\end{figure}

To relieve the overloaded controller, the load balancing framework exploits the fact that multiple coverage areas are overlapping in 5G networks, and mobile users can switch to other networks based on a combination of criteria, using a vertical handover. The criteria used are network, coverage, user, and service based. User preference is taken into account and Analytical Hierarchical Process is used to decompose user preferences, compare them against each other, calculate respective weights for each parameter, and formulate a weight equation to make the handover decision. The authors take into account the fact that user preferences might sometimes be inconsistent, so they check for inconsistencies using a consistency index of the decisions made and comparing it with a random consistency index to find out if the user judgments are inconsistent.  Fig.\ref{loadbalancingalgo} gives an overview of the load balancing method.

\subsubsection{Handover Radio Link Transfer Process}
This is not covered in this paper.

\subsubsection{Impact of SDN and Load-balancing}

To evaluate the proposed framework, Ryu controllers in the Mininet Environment \cite{mininet} is used. Authors show that UDP throughput goes down during handover and jitter goes up, how load balancing tackles this by performing vertical handover and improving the metrics immediately. In another experiment, authors show how their load balancing method affects users’ response time when in high load conditions vs when their method is not used by comparing controller response times. The proposed framework shows a 28\% drop in response time compared to previous proposals. The improvement in performance can be attributed to the load-balancing framework used and the hierarchical architecture of SDN.

Related works covered in this survey talk about improving the handover process by reducing propagation delay, applying thresholding schemes decision before a decision is made using Multi Criteria Decision Making, signaling modifications based on the architecture, etc. This is different from other approaches as a vertical handover between different domains is used as a means of relieving controller signaling load, which in turn reduces the controller's processing delay. Thus, it is useful in alleviating overloaded controllers in situations of dense network traffic thereby improving responsiveness during such peak hours. The only limitation is the fact that RAT-specific cases are not considered while designing this and the question of how the controllers connect to RAT-specific domains is open to interpretation. 

\subsection{SDN based LTE-WiFi specific vertical handover}

The papers discussed in previous subsections have introduced new SDN based architectures that enable multi RAT communication through the use of SDN and proposed ways to use the respective architectures to enable handovers. \cite{liyanage2017sdnoffloading} talks about an SDN based alternative to LTE-WiFI offloading that provides an operator-assisted offloading platform. The 3GPP LTE standard uses a WiFi Offloading approach \cite{ltewifiaggregation},\cite{ltewifiaggregation3gpp}, where data traffic from LTE networks is 'offloaded' onto the WiFi network. This has various benefits such as reducing cellular operator costs, solving network unavailability issues during peak hours, etc. There are some well-known WiFi offloading schemes and some standardized procedures by the 3GPP such as loosely coupled inter networking mechanism S2B, LTE-WiFi IPsec tunnel (LWIP)\cite{lwalwip3gpp}, LTE-WiFi Aggregation (LWA)\cite{ltewifiaggregation3gpp}, and enhanced LWA. \cite{liyanage2017sdnoffloading} explains that these existing offloading schemes are user-triggered, meaning the network conditions are not taken into account as the user has no view of the global network conditions, as mobile applications that the users use have limited access to complete network status indicators. Hence, \cite{liyanage2017sdnoffloading} proposes an operator-assisted offloading platform using SDN, which is not restricted to specific types of RAT, making this a Multi RAT alternative. The advantage of operator-assisted offloading is that the process is initiated by the operator without user intervention, based on accurate real-time network conditions. The paper considers a generic scenario with both single-operator single-controller and multi-operator multi-controller networks. Additionally, the proposed platform is applicable to non-3GPP access technologies like WiFi, LiFi \cite{lifi}, Satellite networks, etc.

\begin{figure}[htbp]
\centerline{\includegraphics[width=\columnwidth]{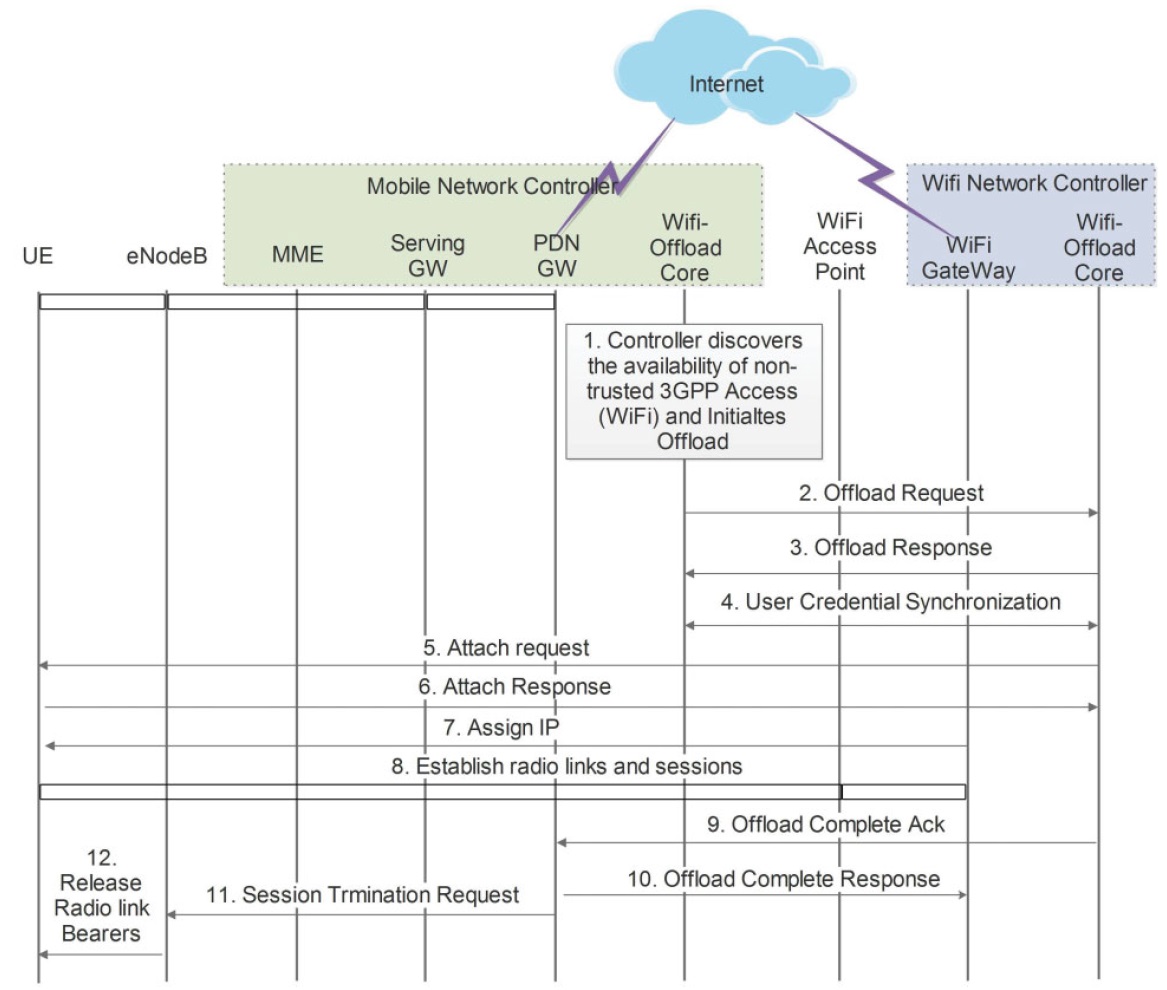}}
\caption{Message exchange between the controllers \cite{liyanage2017sdnoffloading}.}
\label{msgexchangeoffload}
\end{figure}

\subsubsection{Architecture}
The architecture of the offloading platform proposed in \cite{liyanage2017sdnoffloading} is not very informative. The SDN architecture used resembles a centralized architecture, with a controller controlling each RAT. But due to the presence of multiple controllers, this is considered a semi-centralized architecture. The proposed architecture consists of a data offloading core in the SDN controller which houses all the relevant classes and functions required to perform the offloading action. The offloading core is implemented as an application plane program, running on top of each controller. It manages the exchange of user information between multiple SDN controllers. The communication between the controllers only consists of the offloading algorithm, but not any other information such as the synchronization between the two controllers.

\subsubsection{Handover Decision Process}

This paper focuses on how the connection is changed from LTE to WiFi after a decision to handover to the WiFi network is taken. The details about what triggers the handover or offloading are not mentioned.

\subsubsection{Handover Radio Link Transfer Process}

The offloading core in each of the SDN controllers has relevant functions to manage the offloading between WiFi and LTE. It also has a set of modules for three offloading schemes, signal strength based, cost based, and subscriber density based offloading. It is possible to add/drop these schemes according to the offloading algorithms used by the operator. Fig. \ref{msgexchangeoffload} shows the proposed message exchange between different entities to perform the WiFi offloading. The offloading scheme or the handover from LTE to WiFi is triggered when the LTE controller detects the presence of a WiFi network. An offloading request is sent to the offloading core in the WiFi network. If there are enough network resources in the WiFi network to serve the UE, an affirmative offload response is sent to the LTE controller to initiate the handover or the offloading to the WiFi network. The specific network parameters that are considered by the WiFi controller are not mentioned.

\subsubsection{Impact of SDN on the handover}

\begin{figure}[htbp]
\centerline{\includegraphics[width=\columnwidth]{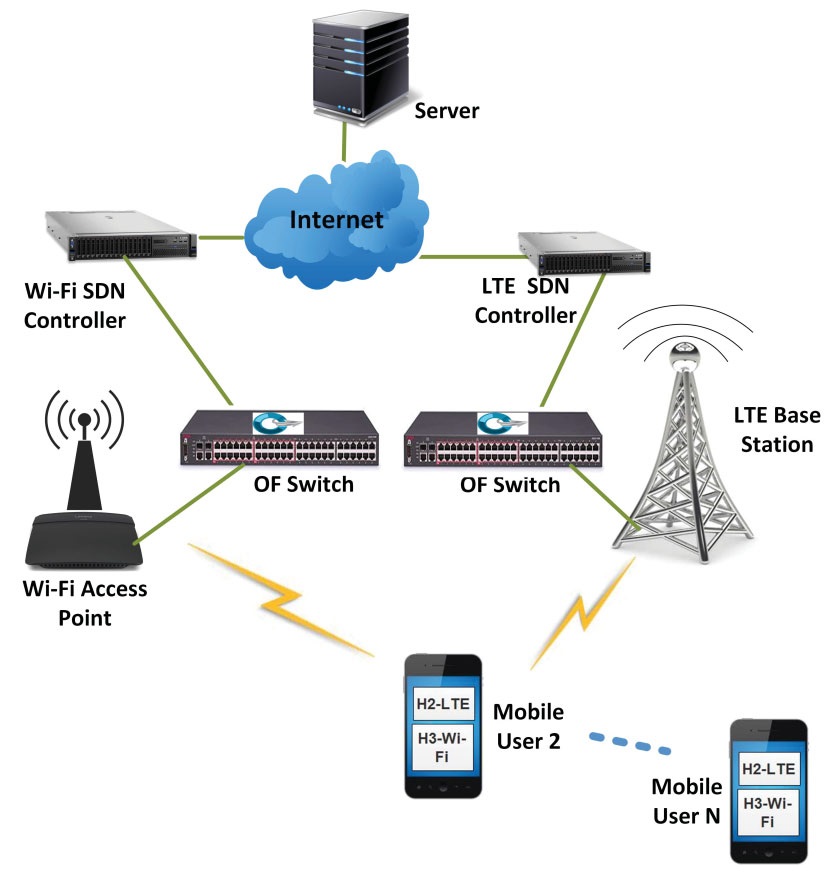}}
\caption{Experiment testbed used in \cite{liyanage2017sdnoffloading}.}
\label{testbedoffloading}
\end{figure}

The platform is implemented on a testbed shown in Fig. \ref{testbedoffloading}. The proof of concept prototype is implemented using Floodlight SDN controller \cite{floodlight}, RESTFul architecture, and the mininet\cite{mininet} platform. The number of users can vary up to 18, each connected to an Iperf server through a client. The authors perform two experiments, one to verify the interoperability of the different offloading mechanisms and another to measure the delay of the offloading platform. 

The experimental results shown in the paper suggest that the proposed mechanism is indeed implementable with decent handover delay performance. The average offloading delay observed was 30.5ms, which was not compared to any other method, such as the delay in the case of the user-initiated LTE-WiFi aggregation that the authors seek to improve. Nevertheless, the paper provides a detailed architecture for operator based offloading, which is relatively new. The ability of SDN to interconnect two different RATs is used to achieve operator-assisted WIFI offloading. But the details of how the Floodlight controllers and OpenFlow switches are interfaced with the LTE base station and WIFI access point respectively are not mentioned, which would have helped researchers implement and verify the performance of the proposed system.

\subsection{SDN based seamless convergence architecture for LTE and WiFi}

\cite{wang2016convergence} proposes an SDN based seamless convergence architecture for seamless handover between LTE and WiFi networks. Authors of \cite{wang2016convergence} explain seamless convergence of WLAN and LTE is challenging due to two reasons: seamless mobility and seamless interoperability between the two networks. They explain that seamless mobility is an issue as WLAN networks are inherently not designed to support mobility, and handover between LTE and WLAN can introduce a longer disruption due to the requirements of re-authentication and service level re-connections. Seamless interoperability demands that a constant quality of service requirement be provided in the different networks. They propose three contributions to address these problems: 

\begin{itemize}
    \item SDN based convergence architecture that separates the control and data planes for LTE and WLAN.
    \item  virtualization based seamless mobility in WLAN.
    \item seamless mobility at a service level for switching between the two networks.
\end{itemize}

\subsubsection{Architecture}

\begin{figure}[htbp]
\centerline{\includegraphics[width=9.25cm]{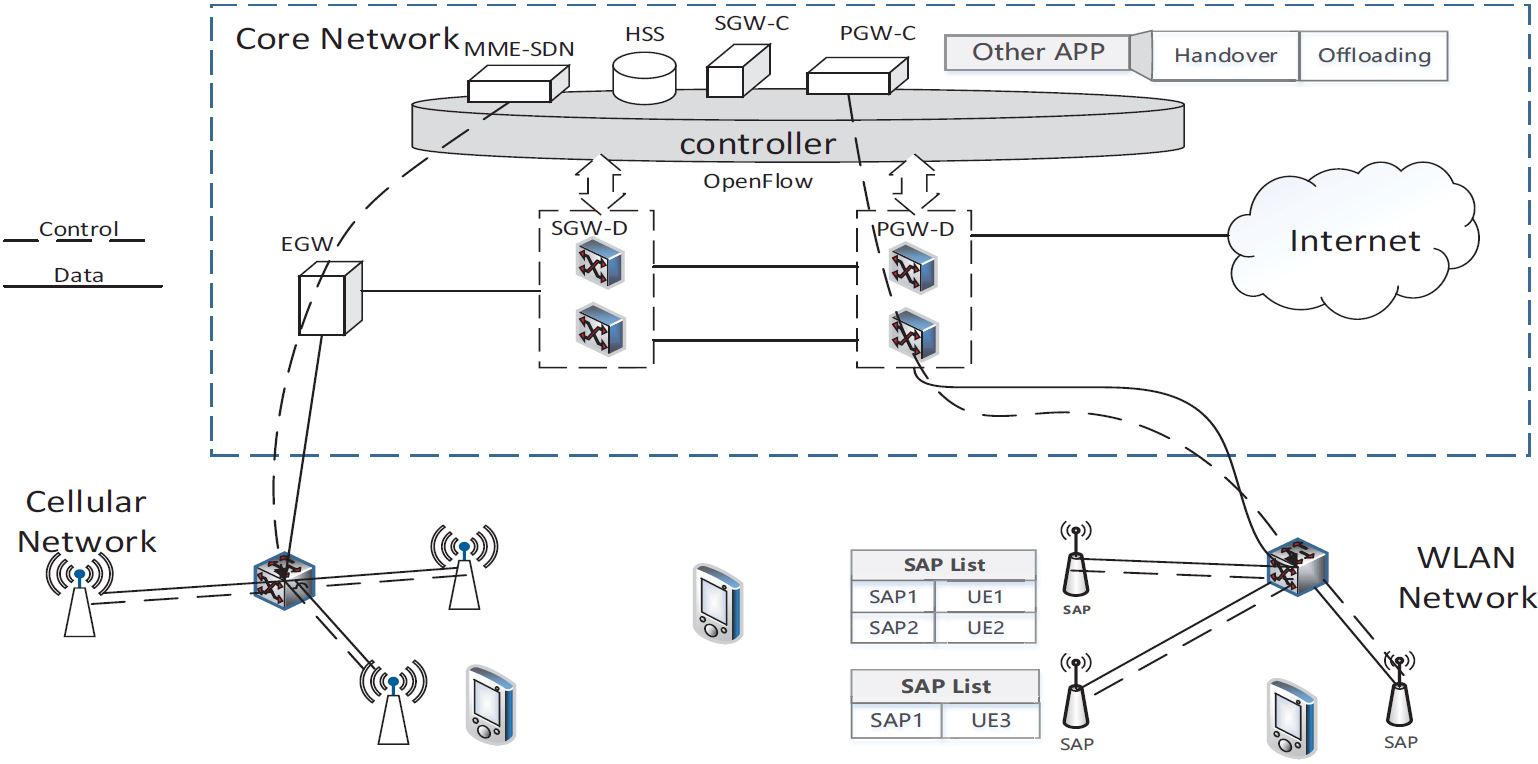}}
\caption{Proposed convergence architecture for LTE and WLAN in \cite{wang2016convergence}.}
\label{convergencearch}
\end{figure}

The architecture in this paper uses a single central SDN controller as shown in Fig. \ref{convergencearch}. The proposed convergence architecture uses SDN Controller to connect with WLAN and LTE after each of the networks is split into a control and data plane respectively. To disaggregate the WLAN, the authors use their previous work called Software defined Wireless Access Network (SWAN)\cite{swanconvergence}, which abstracts the connection between the UE and the WiFi AP using a software defined signal process and implementation, which is termed Soft Access Point(SAP). The Soft Access Point  is essentially a virtual access point holding the layer 2 and layer 3 information of the AP and UE which are used to establish a connection such as the MAC address, AP’s IP address, and UE’s IP address. This way, when a UE connects to the AP, it is assigned a unique Soft Access Point and each physical AP holds a number of Soft Access Point  instances in accordance with the connected UEs. Soft Access Points are retained by the controller regardless of the mobility status of the UE, thereby eliminating the need for multi layer re-connections when the UE moves from one AP to another. This, coupled with the global view of the controller enable an efficient handover. In the LTE network, the PDN-GW and S-GW are decoupled into the user and control plane. 

\subsubsection{Handover Decision Process}
The type of handover used is a vertical handover, but there is no information on how the handover is triggered in the paper.

\subsubsection{Handover Radio Link Transfer Process}

\begin{figure}[htbp]
\centerline{\includegraphics[width=9.25cm]{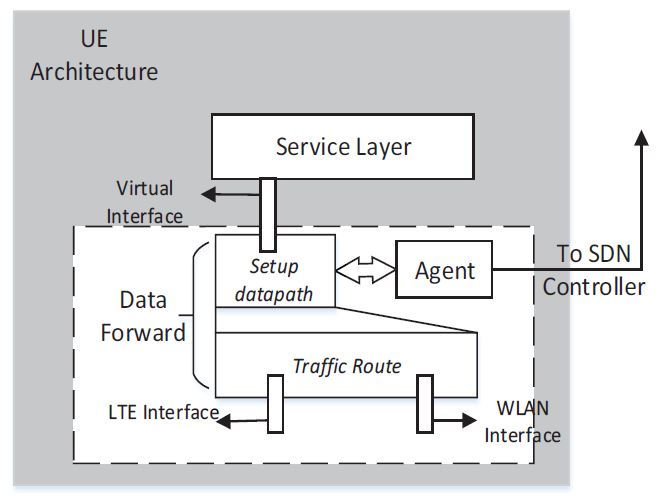}}
\caption{Middle box architecture used in \cite{wang2016convergence}.}
\label{middlebox}
\end{figure}

\cite{wang2016convergence} proposes a virtualization based seamless mobility between LTE and WiFi networks. Large service interruption time is a significant hurdle, which is caused by two reasons. First, the need to change the UE's IP address because of service re-connection to the other network. Second, the core network has to set up different bearers for different access networks, which in the case of a coupled control and data network is a time taking process. To avoid the change in UE’s IP address during the vertical handover between LTE and WiFi, the authors propose to use a virtual middlebox, shown in Fig.\ref{middlebox}. The proposed middle box consists of virtual interfaces, for LTE and WLAN respectively, and an agent and a data forwarding entity. The virtual interfaces create a unique IP address for a UE when it is connected through either LTE or WLAN, which does not change until it leaves the network or becomes idle. So, the need of changing IP address is eliminated which avoids an interruption in services, enabling the seamless handover. 

During normal LTE operation, the middlebox installs flow rules that make the virtual interface forward packets to the LTE interface. Additionally, the core network's flow tables are also configured to transmit data to the LTE network. In the event of a handover trigger, the controller informs the middlebox to connect to WLAN. In the core network, new flow tables are set up by the controller, such that all the packets are copied to the WLAN. Then, the controller instructs the middlebox to change the rules of routing in the UE’s data forwarding module, making the switch to WLAN complete. This way, using SDN's decoupled architecture and by updating flow tables, the second problem of interruption time due to the need to set up different bearers is addressed.

\subsubsection{Impact of SDN based convergence architecture for seamless handover}

The experiments to test the proposed architecture and handover algorithm are conducted on the proposed SDN based WLAN named Software defined Wireless Access Network and an OpenAirInterface (OAI) based LTE platform\cite{oai}. A constant bit rate end-to-end service is implemented and packet delays are measured during the handover phase. results show that using the proposed architecture vertical handover delay is less than 100ms versus a service interruption of more than 3 seconds, seen in the case of a normal LTE to WLAN vertical handover. The improvement in service interruption time can be attributed to the way the SDN based decoupled architecture and flow table updates are used to interface LTE with the WiFi network seamlessly.

\subsection{Observations}

\cite{taksande2020open5g},\cite{wang2017sdn} both build on existing 5G architectures to incorporate inter-RAT operations,  \cite{taksande2020open5g}'s Open5G protocol and the SMRAN architecture provides a comprehensive way of bringing together different RATs. On the other hand, \cite{wang2017sdn} introduces an intriguing Virtual RAT approach, through programmable interfaces and achieving a RAT independent solution of achieving a Multi RAT implementation. While these are not innovations related to mobility management exclusively, they are very good proposals in the domain of Inter-RAT implementations. \cite{alotaibi2018hierarchical} and \cite{alotaibi2021linking} is the converse, concentrating exclusively on the handover optimization in multi RAT networks using a generic SDN based architecture. Further, it is shown by \cite{liyanage2017sdnoffloading} that even existing protocols and procedures like LWA can be improved using SDN. Hence, these works are useful in understanding the benefit of SDN in bringing together different RATs. On exhaustive analysis of these works that cover multi RAT handover optimization, the following are the common shortcomings across these approaches.

\begin{itemize}
    \item There is a lack of a unifying architecture for multi RAT networks, that enables a seamless way of switching between the RATs, as highlighted by \cite{taksande2020open5g} and \cite{wang2017sdn}.
    \item Handover optimization in multi RAT networks and UDNs is a very significant challenge. As the traffic density grows, the problem of redundant and frequent handovers, non-availability of networks increases. These problems are investigated in detail, from various perspectives in the papers, \cite{alotaibi2018hierarchical}, \cite{alotaibi2021linking}. Each of these provides a solution using an SDN based approach of optimizing the handover, by using a threshold for Multi Criteria Decision Making algorithms, predicting the path of the UEs using the SDN's global visibility and designing a new load balancing framework for SDN controllers to consequently reduce the handover processing time respectively. 
    \item The role SDN provides in realizing a unified architecture for multi RAT networks is inspiring. The non-dependency on a specific RAT protocol and the interoperability it can provide as seen in the papers in this section such as in \cite{taksande2020open5g} and \cite{liyanage2017sdnoffloading}, \cite{wang2016convergence} makes the integration of SDN into existing architectures an easy one. Additionally, the unifying global view it offers through hierarchical architectures, as seen in \cite{alotaibi2018hierarchical} enables network operators to have a better view of the network and enables better use of the available RATs, in terms of both HOs between each other RAT and the overall handover performance in individual networks. But the limitation to all these is the fact that there is no simulation platform that can implement the RAT protocols in conjunction with SDN.
\end{itemize}


\section{Open problems and Future work}
\label{futurework}

Throughout the course of the survey, we look at ways in which SDN is applied in various networks and standardized architectures to improve handover and mobility management. This section identifies and summarizes the open problems and provides a list of future works. Additionally, we specifically talk about the implementation platforms, or the lack of these to highlight the emerging need for an all-inclusive simulator that can support multiple RATs and SDN technology.

\subsection{Implementation platforms}

 The available platforms and tools that are available to researchers to implement SDNs, WiFi networks, and cellular networks, individually and combined, can be classified into two broad categories: software based and Hardware.  Hardware implementations are straightforward, by use of OpenFlow enabled switches, Access Points, etc. but there are only a few such implementations seen in the literature. Authors of \cite{liyanage2017sdnoffloading} mention that they implement their offloading architecture using a testbed as shown in Fig.\ref{testbedoffloading}. However, the implementation and interfacing details between the SDN controller and WiFi AP, and the Base station haven't been mentioned. So, most of the implementations covered in existing literature are based on Software tools. An alternative to using software tools is to use analytical and mathematical modeling to verify the proposed solutions.

5G networks consist of many wireless technologies co-existing with them such as LTE, WiFi, LoRaWAN, Bluetooth, and Satellite, all of which are IP based 3GPP and non-3GPP RATs. Due to this coexisting nature, it is of interest to researchers, to simulate these networks in a single universal tool/software that supports all these technologies. Popular choices of researchers are network simulation software such as NS-3\cite{ns3}, OMNET++ \cite{omnet}, GNS3, etc. These are comprehensive network simulators that have been used in wired and wireless networks research but do not support SDN based architectures. NS-3\cite{ns3} provides LENA, and 5G-LENA modules for the simulation of 4G LTE and 5G networks respectively, and the OFSWITCH13 module of NS-3 has support for SDN., but the integration between these different modules is very difficult, and needs a developer based approach. Mininet\cite{mininet} and Mininet WiFi\cite{mininetwifi} are popular tools that serve as network emulators with in-built SDN controllers and OpenFlow enabled switches providing SDN enabled Networks. Mininet WiFi supports SDN-enabled WiFi networks but does not support 5G or LTE. Therefore, there exist very few platforms that can implement SDN based Multi RAT networks. \cite{pham2019enablingsdnadapter} and \cite{Sylla2022Emu5GNet} attempt to solve this problem by bridging these simulators, in such a way, that the proposed integrated tool is capable of supporting heterogeneous SDN based multi RAT networks.

\begin{figure}[htbp]
\centerline{\includegraphics[width=\columnwidth]{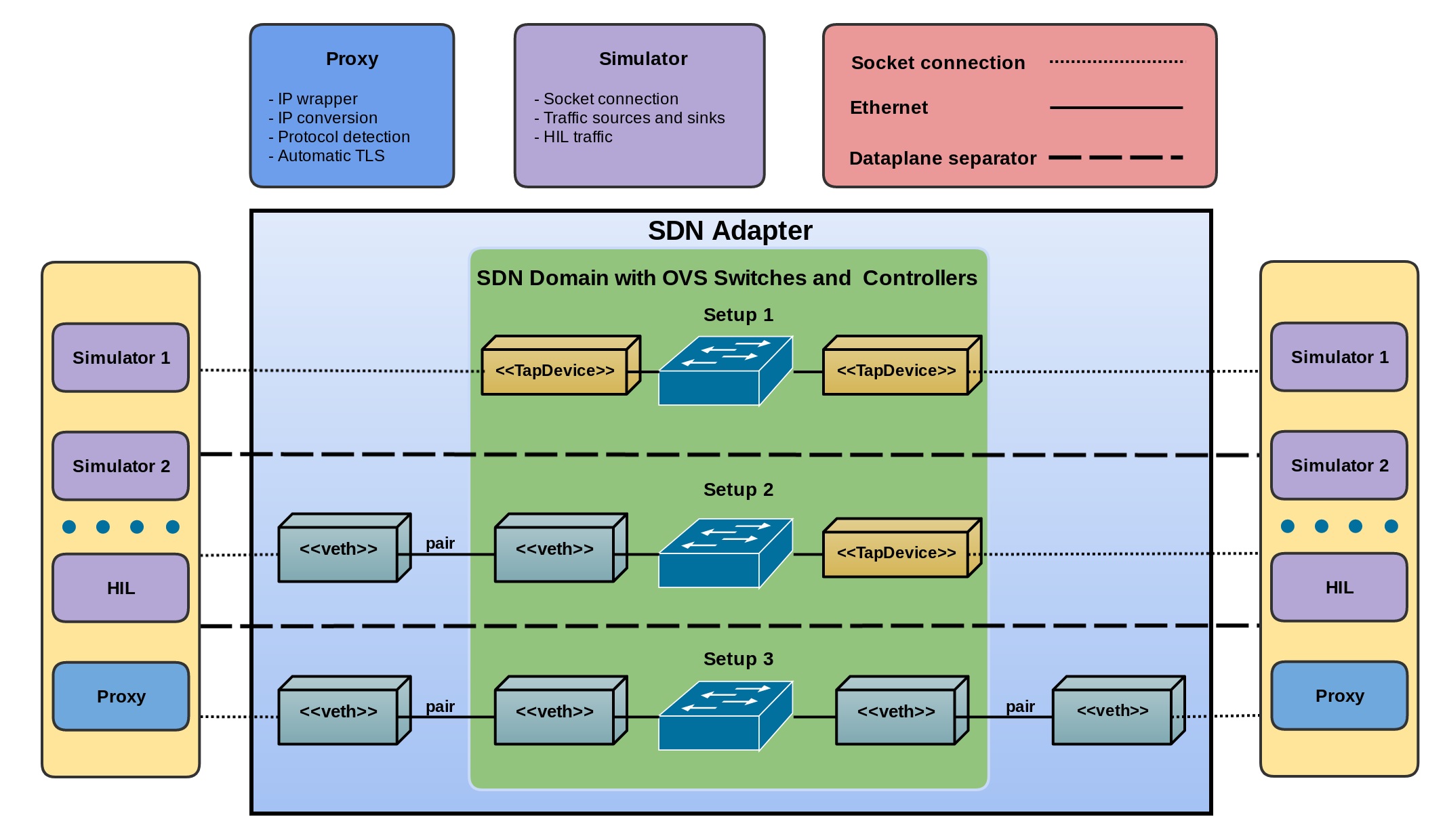}}
\caption{SDN Adapter Architecture by \cite{pham2019enablingsdnadapter}.}
\label{sdnadapter}
\end{figure} 

\begin{figure}[htbp]
\centerline{\includegraphics[width=\columnwidth]{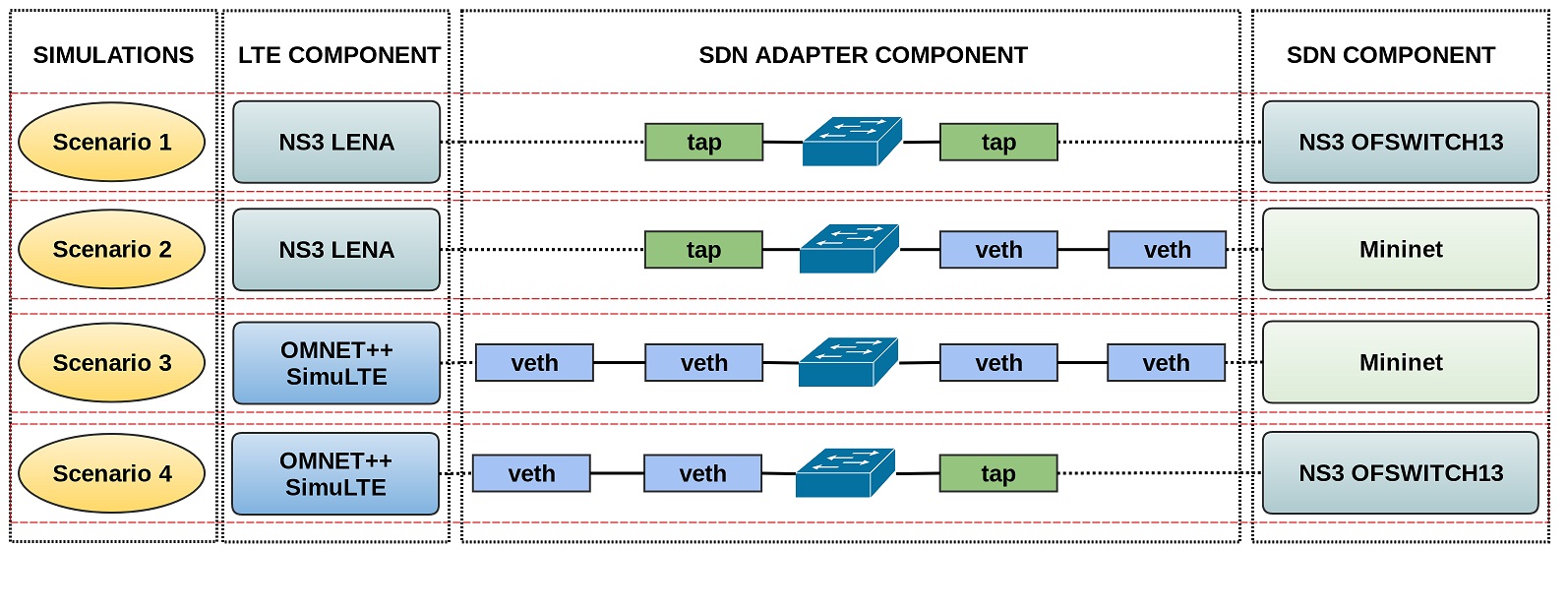}}
\caption{Simulation combinations and their setup as seen in \cite{pham2019enablingsdnadapter}.}
\label{sdnadapterscenarios}
\end{figure}

\cite{pham2019enablingsdnadapter} proposes a tool that seeks to provide interoperability between RATs, by bridging existing software emulators/simulators using an SDN based adapter. The authors develop a novel SDN adapter that interconnects different tools like NS3, Mininet-WiFi, and OMNET++. Fig. \ref{sdnadapter} shows the architecture of the proposed SDN adapter with three main components: SDN, Access Radio, and the SDN Adapter Component. The idea is to interconnect the access radio (LTE/5G) component with an SDN component using the SDN adapter. The SDN adapter connects to each simulation network at the transport layer. It serves as a packet-forwarding device between the different IP based networks it connects to. The control block of the adapter consists of Ryu \cite{ryu} based controllers, that provide Address Resolution Protocol (ARP) responders and Network Address Translation (NAT) service to manage the ingress and egress traffic with OVS flow installation. A proxy block is used to encapsulate the payload into TCP/UDP frames before sending them into the SDN terminal, in the case of legacy protocols, as SDN is inherently designed to only work on IP based networks. The SDN component is the core network that needs to be connected to the access radio, i.e. LTE/5G component, as shown in the figure. For each of these two, two independent simulation platforms are selected and a total of 4 scenarios are implemented as shown in Fig.\ref{sdnadapterscenarios}. LTE component is provided by using NS3 LENA and OMNET++ SimuLTE respectively, while the SDN component is simulated using Mininet WiFi and NS3 OFSWTICH13 respectively. Both the NS3 components are connected to the SDN adapter using an external TapDevice, while OMNET++ and Mininet WiFi are connected using a virtual ethernet (veth) interface. Additionally, to enable the easy set-up of the mentioned 5G simulation scenarios, an SDN Adapter Designer GUI tool is provided, enabling users to intuitively configure the adapter configuration to suit each of the four scenarios shown in Fig.\ref{sdnadapterscenarios}. The authors also extend the simulation scenarios to incorporate 5G networks, by demonstrating this by simulating a 5G network slice with OAI5G, NextEPC, SDN Adapter, and Mininet-WiFi.

\cite{Sylla2022Emu5GNet} proposes Emu5GNet, a tool that provides support for the emulation of multiple RATs (5G and WiFi) and their corresponding core networks. It provides a simulation platform that supports different types of mobile traffic and the deployment of edge servers with an SDN/NFV architecture. Emu5GNet integrates the Mininet WiFi based Containernet with the VIM-EMU, UERANSIM and Open5GC, to obtain a platform that can implement an SDN/NFV based end-to-end 5G and WiFi networks. Containernet is an extension of Mininet WiFi, with Docker based containers serving as hosts, and providing emulation of SDN based WiFi networks. The VIM-EMU platform, which is also based on Docker containers, is a point-to-point NFV emulation platform that provides an efficient way to deploy and manage NFV functions and Edge services in wired networks. It facilitates the development and testing of network services in realistic end-to-end multi point-to-point scenarios. The authors propose to extend this functionality to wireless 5G networks and to integrate VIM-EMU with Containernet, such that edge data centers can be deployed using VIM-EMU in the topologies and architecture offered by the Mininet WiFi based Containernet. Next, UERANSIM is a tool that provides an implementation of 5G UEs and RAN, in accordance with 3GPP's standalone and non-standalone architectures. Open5GS is a tool that offers a complete 5G core implementation, designed to interconnect with 5G RAN platforms, such as the UERANSIM. The authors integrate Open5GS and UERANSIM, into the Mininet WiFi containernet, to allow for a large-scale end to end deployment of 5G networks. Therefore, this bridges Mininet WiFi's support for the WiFi networks  with an end-to-end deployment of the 5G RAN and CN using UERANSIM and Open5GS respectively. 

\begin{figure}[htbp]
\centerline{\includegraphics[width=\columnwidth]{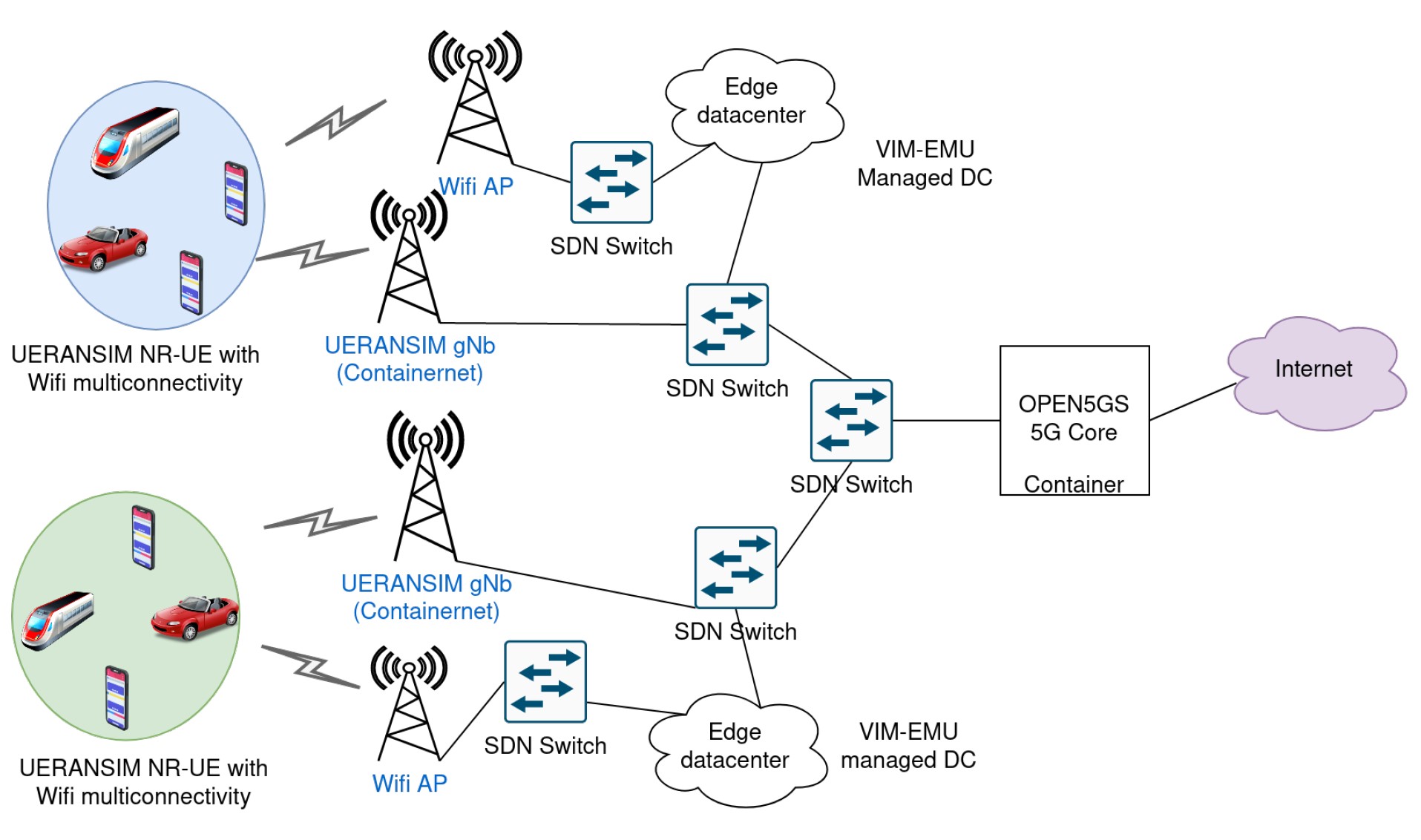}}
\caption{Architecture of Emu5GNet, as seen in \cite{Sylla2022Emu5GNet}.}
\label{emu5GNet_architecture}
\end{figure}

The architecture of the proposed Emu5GNet is shown in Fig.\ref{emu5GNet_architecture}. Emu5GNet provides a robust way of implementing multi RAT networks using SDN and NFV based architectures. The architecture proposed supports complete 5G protocol stacks and SDN based switches that connect WiFi and 5G nodes. While technical details of the integration between the different tools used in the Emu5GNet are not specified, it is mentioned that all the components use Docker containers, and the non-docker based tools such as UERANSIM and Open5GS were dockerized and integrated with containernet. So, this makes it a Docker based integration. Additionally, as this is a tool in its nascent stages of deployment, its reproducibility could not be tested.

\cite{pham2019enablingsdnadapter} provides an innovative way of solving the crucial problem of interconnecting Heterogeneous SDN enabled 5G networks through the proposed SDN adapter. While the paper and the proposed SDN adapter seem promising, recreation attempts were unsuccessful. So the reproducibility of the proposed SDN adapter remains a challenge. Along similar lines, \cite{Sylla2022Emu5GNet} also proposes to interface various tools that offer 5G, WiFi, NFV, and SDN capabilities using the proposed Emu5GNet. As the technical aspects of integration are not mentioned, it is not viable to compare the Emu5GNet with the SDN adapter based solution. The SDN adapter based integration of \cite{pham2019enablingsdnadapter} relies on using an SDN element to bridge different components using TapDevice and virtual ethernet (veth) devices, at the level of the transport layer. While the Emu5GNet uses a Docker based integration. The technical details about the integration methods used in Emu5GNet are not known. Furthermore, the Emu5GNet platform could not be tested, due to the various installation requirements. Therefore, from an implementation standpoint, there is no single tested tool/platform that can implement SDN based Multi RAT networks. While \cite{pham2019enablingsdnadapter} and \cite{Sylla2022Emu5GNet} propose tools that can address these, the reproducibility of the tools and experiments conducted in these papers remain untested.

\subsection{SDN and 5G Interaction}

There is a limited amount of research in applying SDN to isolated 5G networks, as seen in section \ref{5gsection}. Theoretically, incorporating SDN into 5G networks is expected to improve the handover processing time as the amount of control signaling messages are reduced. A general approach as seen in \cite{ciciouglu2021multi} and \cite{abdulghaffar2021modeling} is to use the SDN to design and modify the existing Xn based call flow or improve the decision making process of the handover. However, these works fail to comprehensively compare with the existing Xn based protocol. The approach of using an SDN in the 5G architecture to improve handover needs to be comprehensive and be able to support different kinds of network traffic that 5G supports. Additionally, as seen in \cite{dhruvik2021design}, a lack of a 5G protocol stack in the simulation platforms of the NS3 mmWave module limits the way researchers can perform such comparisons with the standards. So, an SDN based handover optimization incorporated with the existing 5G standards combined with implementation on a testbed that supports the SDN paradigm and the 5G NR protocol stack is needed.

\subsection{SDN based Handover Optimization}
Through the various schemes of handover and mobility management covered in this paper, across sections \ref{ltesection},\ref{5gsection},\ref{multiratsection} there are many possibilities of combining complementary optimization approaches. Works such as \cite{gharsallah2019sdn}, and \cite{wang2017sdn} propose software-controlled handover and configuration of the RAN, making the RAN architectures more dynamic and flexible. 

\subsection{Multi Radio Access Network (RAT) Handover Architectures}

An interesting direction is to adopt such architectures that support MultiRAT with novel handover schemes like the, \cite{ciciouglu2021multi}. For example, \cite{alotaibi2018hierarchical} adopts a similar concept to \cite{wang2017sdn}, with respect to estimating the path a UE might take. While the former uses a fixed hierarchical architecture, the latter uses an architecture that is virtualized and provides programmable interfaces that can be tuned to any hardware implementation, allowing a MultiRAT architecture, It would be interesting to combine the two approaches, with the virtualized RAT of \cite{wang2017sdn} and the hierarchical nature of \cite{alotaibi2018hierarchical}'s architecture to estimate the path a UE follows and explore the handover performance between the different possible RATs.

\subsection{Operator-initiated Multi RAT Handover}
As mentioned above, due to the advent of 5G networks, there is a vast co-existence of IP based networks with the 5G networks, such as LTE, WiFi, IoT, D2D, Satellite, etc. \cite{liyanage2017sdnoffloading} shows how operator-initiated LTE WiFi aggregation can be achieved using SDN. At the core of the proposal in \cite{liyanage2017sdnoffloading} is the use of SDN to connect the LTE eNB and WiFI AP, instead of the usual connection seen in the standards. The advantage as highlighted is that the hand-off/offloading is operator initiated and thus has a better view of the network conditions and can be fine-tuned to suit the conditions. Along similar lines, \cite{wang2016convergence} proposes a convergence of WiFi and LTE using SDN. They also propose a seamless vertical handover among the two networks using the middlebox and SDN. Many research directions stem from this approach. First, this idea of offloading can be extended to the two-way inter-connection between cellular and WiFi networks, enabling seamless two-way communication between the RATs and movement of data packets between the two networks, as seen in \cite{wang2016convergence}.  Additionally, the convergence architecture and vertical handover between the RATs seen in \cite{wang2016convergence} can easily be extended to other RATs as  well, owing to the compatibility of SDN with IP based networks. In such a case, the UEs can be offloaded or handed over to the best available RAT network, not just to WiFi. 

\subsection{Mobility Management Algorithms}
An overlooked direction in mobility management is consolidating the research in different aspects of mobility management and handover design. For example, the algorithms and approaches proposed in \cite{ciciouglu2021multi}, look at how applying context awareness, thresholds, and Multi Criteria Decision Making techniques improves the handover makes them seamless and efficient, while the implementations in  \cite{alotaibi2021linking} and concentrate on reducing the controller load and reducing the message processing delay at the controller by designing a new buffer. Each of these looks at different aspects within the bigger problem of improving mobility management, but bringing each of these approaches together can yield a streamlined, efficient, scalable, and seamless handover.

\section{Conclusion}
\label{conclusion}

This paper provides a survey of the SDN enabled architectures and handover optimization techniques in multi RAT networks. Specifically, the improvement SDN brings to existing networks and handover techniques, and how SDN can enable communication between different radio access technologies. As discussed in the individual sections, the direction that the evolution of SDN enabled RAN is taking is an interesting one with a multitude of possibilities. Specifically, due to the nature of this domain, there is not a single tool/software that makes the implementation of the multi RAT heterogeneous networks possible, as outlined in section \ref{futurework}. We cover the potential open directions stemming from these papers in section \ref{futurework} to give the reader an idea of how the existing methods in multi RAT networks can be improved.

\section*{Acknowledgment}
This material is based upon work supported by the National Science Foundation under Grant Number 2030122.

\bibliographystyle{plain}
\bibliography{sdnhandover}

\end{document}